\documentclass[a4paper,twocolumn,numberedappendix,twocolappendix,revtex4,iop]{openjournal}

\usepackage[T1]{fontenc}
\usepackage[utf8]{inputenc}
\usepackage{natbib}

\usepackage{amsmath,amssymb}
\usepackage[a4paper,marginpar=10pt,headheight=20pt, right=0.5cm, top=1.5in, left=2cm, bottom=0pt, footskip=0pt, footnotesep=10pt]{geometry}
\usepackage[normalem]{ulem}
\usepackage{enumitem}

\usepackage{booktabs}
\usepackage{graphicx}	
\usepackage{bm}		
\usepackage{pdflscape}	
\usepackage{subfigure}
\usepackage[usenames,dvipsnames]{xcolor}
\usepackage[colorlinks,linkcolor=blue,citecolor=linkorange,urlcolor=blue ]{hyperref}
\usepackage{multirow}
\usepackage{orcidlink}
\usepackage{float}
\usepackage{url}

\definecolor{linkorange}{rgb}{0.7,0.1,0.0}
\definecolor{valecol}{rgb}{0,0.5, 1.}

\newcommand{\pinocchio}{\texttt{\small PINOCCHIO}} %
\newcommand{\glamer}{\texttt{\small GLAMER}}

\newcommand\dd{\textrm{d}}
\newcommand{\Msun}{M_{\odot}}
\def\d{{\rm d}}

\def\apjl{ApJ Let.}

\def\aap{A\&A}

\def\apjs{ApJS}

\providecommand{\eprint}[1]{\href{http://arxiv.org/abs/#1}{#1}}
\providecommand{\adsurl}[1]{\href{#1}{ADS}}

\usepackage{ifthen}\def\eprinttmp@#1arXiv:#2 [#3]#4@{\ifthenelse{\equal{#3}{x}}{\href{http://arxiv.org/abs/#1}{#1}}{\href{http://arxiv.org/abs/#2}{arXiv:#2} [#3]}}
\renewcommand{\eprint}[1]{\eprinttmp@#1arXiv: [x]@}

\begin{document}

\title{A deconstruction of methods to derive one-point lensing statistics}

\author{
    Viviane Alfradique,$^{1,2}$\footnote{These authors contributed equally to this work.}\orcidlink{0000-0002-5225-1923}
    Tiago Castro,$^{3,4,5,6}$\footnotemark[*]\orcidlink{0000-0002-6292-3228}
    Valerio Marra,$^{3,4,7}$\orcidlink{0000-0002-7773-1579}
    Miguel Quartin,$^{1,8,9}$\orcidlink{0000-0001-5853-6164}\\
    Carlo Giocoli$^{10,11}$\orcidlink{0000-0002-9590-7961}
    and Pierluigi Monaco$^{3,4,6,12}$\orcidlink{0000-0003-2083-7564}}
\affiliation{
    $^{1}$Instituto de Física, Universidade Federal do Rio de Janeiro, 21941-972, Rio de Janeiro, RJ, Brazil\\
    $^2$Centro Brasileiro de Pesquisas F\'isicas, Rua Dr. Xavier Sigaud 150, 22290-180 Rio de Janeiro, RJ, Brazil\\
    $^{3}$INAF – Osservatorio Astronomico di Trieste, via Tiepolo 11, I-34131 Trieste, Italy\\
    $^{4}$INFN – Sezione di Trieste, I-34100 Trieste, Italy\\
    $^{5}$IFPU -- Institute for Fundamental Physics of the Universe, via Beirut 2, 34151, Trieste, Italy\\
    $^{6}$ICSC - Centro Nazionale di Ricerca in High Performance Computing,
    Big Data e Quantum Computing, Via Magnanelli 2, Bologna, Italy\\
    $^{7}$Departamento de Física, Universidade Federal do Espírito Santo, 29075-910, Vitória, ES, Brazil \\
    $^{8}${PPGCosmo, Universidade Federal do Espírito Santo, 29075-910, Vitória, ES, Brazil}\\
    $^{9}$Observatório do Valongo, Universidade Federal do Rio de Janeiro, 20080-090, Rio de Janeiro, RJ, Brazil\\
    $^{10}$INAF - Osservatorio Astronomico di Bologna, via Piero Gobetti 93/3, I-40129 Bologna, Italy\\
    $^{11}$INFN, Sezione di Bologna, viale Berti Pichat 6/2, I-40127 Bologna, Italy\\
    $^{12}$Università di Trieste, Dipartimento di Fisica, via Tiepolo 11, 34143 Trieste
    }


\begin{abstract}
Gravitational lensing is a crucial tool for exploring cosmic phenomena, providing insights into galaxy clustering, dark matter, and dark energy. Given the substantial computational demands of $N$-body simulations, approximate methods like \pinocchio\ and \texttt{turboGL} have been proposed as viable alternatives for simulating lensing probability density functions (PDFs). This paper evaluates these methods and their effectiveness across both weak and strong lensing regimes, with a focus in the context where baryonic effects are negligible. Our comparative analysis reveals that these methods are effective for applications where lensing is mild, such as the majority of sources of electromagnetic and gravitational waves. However, both \pinocchio\ and \texttt{turboGL} break down for large values of convergence and magnification due to their loss of accuracy in capturing small-scale nonlinear matter fields, owing to oversimplified assumptions about internal halo structures and reliance on perturbation theory. \pinocchio\ yields second-to-fourth moments of the lensing PDFs, which are 6--10\% smaller than those resulting from $N$-body simulations in regimes where baryonic effects are minimal.
These findings aim to inform future studies on gravitational lensing of point sources, which are increasingly relevant with upcoming supernova and gravitational wave datasets.
\end{abstract}

\keywords{large-scale structure of Universe -- cosmology:observations -- cosmological parameters}

\maketitle

\section{Introduction}\label{sec:intro}

Electromagnetic and gravitational wave (GW) point sources, such as supernovae (SN) and compact binary coalescences, have proven to be invaluable for studying the universe's evolution across extensive redshift ranges. Upcoming observations such as those from the Euclid mission~\citep{Euclid:2024yrr},\footnote{\url{https://www.euclid-ec.org/}} the Nancy Grace Roman Space Telescope,\footnote{\url{https://roman.gsfc.nasa.gov/}} and the Vera C. Rubin Observatory\footnote{\url{https://www.lsst.org/}} are set to accurately observe tens of thousands of supernovae up to a redshift of $z\sim2$. Simultaneously, third-generation gravitational wave detectors such as the Einstein Telescope\footnote{\url{https://www.et-gw.eu}} and Cosmic Explorer\footnote{\url{https://cosmicexplorer.org}} will detect hundreds of thousands of gravitational events up to $z\sim5$ and beyond. To harness this wealth of data effectively, it is crucial to meticulously model all potential systematic effects, with gravitational lensing being a significant concern, particularly for sources at $z \gtrsim 1$, as it strongly influences distance measurements.

Gravitational lensing affects electromagnetic and gravitational waves as they traverse our inhomogeneous Universe. This phenomenon describes the deflections that waves undergo when interacting with the matter density field, causing distant sources to appear magnified or demagnified. Such deflections can lead to significant distance measurement errors, modifying the intrinsic distribution of astronomical sources and introducing variance and non-Gaussianities. Furthermore, these deflections also give rise to distortions in the shapes of the background galaxies. A gravitational lens induces a correlation on galaxy image orientations, which can be measured statically and reveal the properties and evolution of large-scale structures. This is known as \textit{cosmic shear} \cite[see, e.g.][for a review]{Bartelmann2001,Kilbinger:2014cea}.

Gravitational lensing notably limits the accuracy of distant standard candles and sirens as cosmological probes. Estimates suggest that lensing adds a relative uncertainty of 0.5\%--1\% to distance measurements of stellar binary black holes up to 3 Gpc~\citep{Shan2021}. Recent studies, such as those by \citet{Mpetha2024}, have indicated that neglecting this effect can bias cosmological analyses of bright sirens at $z\leq0.5$ by more than 1$\sigma$, resulting in $\Delta H_{0} = \pm$ 0.25 km/s/Mpc for binary neutron stars (BNS) and $\Delta H_0 = \pm$ 5 km/s/Mpc for supermassive black hole binaries. \citet{Canevarolo:2023dkh} further demonstrated that gravitational lensing could introduce biases up to 3.5$\sigma$ for the $\Omega_{\rm m0}$ measurement and up to 2.5$\sigma$ for $H_0$ using BNS up to $z=2$. Lensing may also bias the distribution of neutron star masses~\citep{Canevarolo:2024muf}. Moreover, \citet{Wu2023} showed that `delensing' methods, which correct the effect after estimating the lensing magnification independently through weak lensing maps reconstructed from galaxy surveys, could halve these lensing errors on average \citep{Shapiro2010}.

The lensing probability distribution function (PDF) is the fundamental one-point statistic for describing the lensing of point sources.
After some pioneering work (e.g.~\citealt{Frieman:1996xk}), various approaches have been followed so as to compute the lensing PDF relative
to a given cosmology. A first approach (e.g.~\citealt{Munshi:1999qw, Valageas:1999ir, Wang:2002qc,Das:2005yb}) relates a ``universal'' form of the lensing PDF to the variance of the convergence, which is fixed by the amplitude $\sigma_{8}$ of the power spectrum. Moreover the coefficients of the proposed PDF may be trained on a grid of $N$-body simulations.
Recently, using theoretical predictions from Large Deviation Theory to incorporate the non-Gaussian effects of the lensing PDF, \cite{Barthelemy2020} showed that it is possible to compute the lensing PDF consistently with results from simulations at $z=1$ and for an opening angle of 10 arcmin.
Alternatively, one can build a model of the universe, e.g., using an $N$-body simulation, and directly compute the relative lensing PDF, usually through time-consuming ray-tracing techniques \citep[e.g.,][]{Holz:1997ic,Bergstrom:1999xh, Holz:2004xx,Hilbert:2007ny, Hilbert:2007jd, Takahashi:2011qd,Bolejko:2012ue}. The flexibility of this method is
therefore penalized by the increased computational time.

The most direct and precise method for determining the lensing PDF involves applying the ray-tracing technique in full hydrodynamical cosmological simulations \citep{Castro2018}. However, these simulations demand even more computational resources than their $N$-body counterpart. In regimes with negligible baryonic effects on lensing, $N$-body simulations using collisionless particles provide a sufficient alternative and significantly reduce computational demands. Despite the reliability of the results from $N$-body simulations \citep{Takahashi2011, Hilbert2019}, the computational requirements and storage capacities remain challenging. An even faster approach, which searches for an optimum compromise between precision and speed, is to use approximate methods to describe the nonlinear evolution of perturbations \citep{Manera2013, Kitaura2013, Tassev2013,Monaco:2016pys}. The \textit{PINpointing Orbit Crossing Collapsed HIerarchical Objects} (\pinocchio) algorithm \citep{Monaco2002, Taffoni2002, Munari2017} integrates Lagrangian perturbation theory (LPT) with a semi-analytical model to approximate the nonlinear hierarchical formation of structures. This method generates dark matter halo catalogs using a fraction of the resources needed for $N$-body simulations. It achieves roughly 10\% accuracy in clustering statistics (power spectrum, bispectrum, and two-point correlation function) compared to the $N$-body solution. For the weak-lensing regime, the stochastic gravitational lensing (sGL) approach provides an even faster alternative \citep{Kainulainen2009, Kainulainen2011a, Kainulainen2011b}. This method calculates the 1-halo term of the lensing PDF using stochastic configurations of inhomogeneities based on a halo model and subsequently incorporates the 2-halo term through convolution. A similar analytical method, also based on combining the 1-halo with the 2-halo term, was recently proposed by \citet{Thiele:2020rig}.

The main importance of computing the lensing PDFs is understanding the extra scatter on future SN and GW measurements: an accurate modeling can yield significant cosmological insights. Since the final observed scatter on a standard candle Hubble diagram is a convolution of their intrinsic distribution with the lensing PDF at the observed redshift, one can transform what appears as noise in Hubble diagram residuals into a valuable signal. Numerous studies~\citep{Quartin2014, Castro:2014oja, Castro:2015rrx, Castro:2016jmw, Macaulay:2016uwy, Scovacricchi:2016ylt,DES:2020kbf,DES:2024lto} have investigated the correlation between these residuals and lensing signals, examining how these relationships depend on the background cosmology and the perturbations in the intervening matter. These efforts have demonstrated the potential to achieve a precision of 0.6\% for the power spectrum amplitude $\sigma_8$ and 7\% for the growth rate index~$\gamma$ in forthcoming surveys \citep{Amendola2015}.  One of the proposed methods to constrain cosmological parameters through the Hubble diagram residuals is the \emph{method of the moments} (MeMo, see~\citealt{Quartin2014}).%
\footnote{See also \citet{Patton:2016umg,Liu:2018dsw,Euclid:2023uha,Castiblanco:2024xnd}.}
The MeMo relies on a comparison of the observed central moments of the residual magnitude distribution with the expectations from lensing simulations. \cite{Marra:2013roi} computed fitting functions of the moments, but relied on the weak-lensing based \texttt{turboGL} and only for moderate redshifts. The next generations of SN surveys and GW detections will extend the observed redshift range, and to access all the cosmological information contained in these data more accurate tools that correctly describe the matter distribution in the universe are required.
More recently, \cite{Boyle2021} showed that the lensing PDF provides, when combined with a Planck prior, more powerful constraints on the dark energy equation of state and neutrino mass than the two-point correlation function.

In this paper, we deconstruct the essential physics needed to calculate lensing PDFs accurately. We explore the effect of the 2-halo term, the influence of non-collapsed matter in the field, the accuracy of \pinocchio, and the degree to which baryonic effects can be overlooked. We specifically analyze four relevant lensing quantities in one-point statistics: the PDFs for convergence ($\kappa$), shear ($\gamma$), magnification ($\mu$), and magnitude magnification ($\Delta m$), the latter being particularly crucial for the statistics of lensing of standard candles like SN and GW. We will describe our methodology in Section~\ref{sec:lenssims}, present our comparative results in Section~\ref{sec:test}, and summarize our findings in Section~\ref{sec:concl}. In Appendix~\ref{app:resolutions}, we revise the impact of the angular resolution on the lensing PDF.

\bigskip
\section{Methodology}
\label{sec:lenssims}

\setlength\tabcolsep{3pt}
\renewcommand{\arraystretch}{2}
\begin{table}
\caption{Simulation methods ordered according to their complexity and the range of physical phenomena modeled.}
\label{tab:summary}
\centering
\begin{tabular}{l|p{6.5cm}}
\hline
\hline
model name & description \\
\hline
\texttt{tgl} 1h       &  Stochastic gravitational lensing simulation with the 1-halo turboGL code.\\
\texttt{tgl} 1h+2h    &  Extends \texttt{tgl} 1h by including a theoretical 2-halo term convolution.\\
\pinocchio${}_{\text{halos}}$ &  Ray-tracing on \pinocchio\ past light cone (PLC) with field particles randomly redistributed.\\
\pinocchio    &  Ray-tracing on standard \pinocchio\ PLC.\\
$N$-body       &  Ray-tracing on $N$-body simulation PLC.\\
Hydro        &  Ray-tracing on hydrodynamical simulation PLC.\\
\hline
\end{tabular}
\medskip
\end{table}

This study aims to assess the accuracy of various simulation methods in deriving the lensing Probability Density Function (PDF) on the regime in which baryonic physics is sub-dominant. The methods under investigation are summarized in Table~\ref{tab:summary}, ranked by their complexity and the breadth of physical phenomena they encompass. For the \pinocchio\ and $N$-body simulations, we further evaluate the angular power spectrum and the generated lensing and density maps. The Hydro simulations are used to define the boundaries of reliability of the $N$-body results. For all methods but the Hydro, our fiducial model for the Universe is the $\Lambda$CDM model with the cosmological parameters set very close to the ones obtained by the combination of DES Y1 + Planck 2015 + JLA SNe + 
BAO~\citep{DES:2017myr}: $\Omega_{\rm m0}=0.301, \, \Omega_{\rm b0}=0.048, \, \Omega_{\Lambda 0}=0.699, \, h=0.682, \, \sigma_{8}=0.798$, and $n_{\rm s}=0.973$. For the Hydro simulations, we rely on the lensing PDFs from~\citet{Castro2018}, based on the \textit{Magneticum} simulations.%
\footnote{\url{http://www.magneticum.org/}} \textit{Magneticum} assumes the following background cosmology: $\Omega_{\rm m0}=0.272, \, \Omega_{\rm b0}=0.048, \, \Omega_{\Lambda 0}=0.728, \, h=0.704, \, \sigma_{8}=0.809$, and $n_{\rm s}=0.963$.
We proceed to describe each method in detail in the order they appear in Table~\ref{tab:summary}.

\subsection{The sGL method}

The turboGL (\texttt{tgl}) code\footnote{\url{https://github.com/valerio-marra/turboGL}} is the numerical implementation of the semi-analytical stochastic gravitational lensing (sGL) method, which was introduced in \citet{Kainulainen2009,Kainulainen2011a}.
The sGL method operates by modeling the matter density contrast according to the halo model (HM), where the inhomogeneous Universe is approximated as a collection of different types of halos whose positions satisfy the linear power spectrum.

\subsubsection{The halo model}

The halo model assumes that on small scales (large wavenumbers~$k$) the statistics of matter correlations are dominated by the internal halo density profiles. In contrast, on large scales, the halos are assumed to cluster according to linear theory. The model does not include intermediate low-density structures such as filaments and walls.
Following \citet{Kainulainen2011b}, in \texttt{tgl} the total power spectrum is obtained by the simple addition of the two components:\footnote{More sophisticated and accurate versions of the HM are available in the literature; see \citet{Asgari2023mej}, and references therein.}
\begin{equation} \label{pktot}
P_m(k,z) = P_{L}(k,z) + P_{H}(k,z) \,.
\end{equation}
The first term on the right-hand side is also called the 2-halo component, and the second is the 1-halo component. The usefulness of the halo model stems from the fact that both terms in Eq.~(\ref{pktot}) can be computed without having to resort to numerical simulations. Specifically \citep{Peacock:2000qk}:
\begin{align}
P_{L}(k, z) & = {2 \pi^{2} \over k^{3}}   \delta_{H0}^{2}  \left( {c k \over H_{0}} \right )^{3+n_{s}} T^{2}(k) D^{2}(z) \,, \label{PkL} \\
P_{H}(k,z) &=  \int_{M_{\rm min}}^{\infty} {\rm d}n(M,z)  \left({M \, W_k(M,z) \over \rho_{M0}} \right)^{2}   \,, \label{PkH}
\end{align}
where $T(k)$ is the transfer function, $D(z)$ the growth function, $H_{0}$ is the present-day Hubble parameter, $n_{s}$ is the spectral index, and $\delta_{H0}$ is the amplitude of perturbations on the horizon scale today, which we fix by setting $\sigma_{8}$.
Furthermore, in Eq.~\eqref{PkH}, $M_{\rm min}$ is the smallest halo mass that is considered, $\d n$ is the number density of halos in the mass range $\d M$, which is defined via the halo mass function $f(M,z)$ that gives the fraction of the total mass in halos of mass $M$ at the redshift $z$:
\begin{equation}
\d n(M,z) \equiv n(M,z) \d M = {\rho_{M0} \over M} \, f(M,z) \d M \, ,
\label{eq:hmf}
\end{equation}
where  $n(M,z)$ is the number density and $\rho_{M0}$  the matter density today.
Finally, $W_k$ is the Fourier transform of the halo density profile:
\begin{equation}
W_k(M,z)=  \frac{1}{M} \int_{0}^{R} \rho(r,M,z)  {\sin k r \over k r}  4\pi \, r^2 \, {\rm d}r \,,
\end{equation}
and $R$ is the halo radius. For the halo profile $\rho(r,M,z)$, we use the Navarro-Frenk-White (NFW) profile~\citep{Navarro:1995iw}.
Modifications to the inner cuspy NFW profile by baryonic feedback could be included if a suitable profile is provided \citep[see, e.g.,][]{DiCintio:2013qxa,Schneider:2015wta}.

\subsubsection{Lensing convergence}

The lens convergence $\kappa$ in the Born approximation is given by the following integral evaluated along the unperturbed light path~\citep{Bartelmann2001}:
\begin{equation} \label{eq:kappa1}
    \kappa(z_{s})=\int_{0}^{r_{s}}\dd r \, \rho_{M0} \, G(r,r_{s})\,\delta_{M}\big(r,t(r)\big) \,,
\end{equation}
where $G$ is the lensing efficiency of inhomogeneity at the comoving radius $r$ (basically the inverse of the critical surface density defined below):
\begin{equation} \label{opw}
G(r,r_{s})=  \frac{4\pi G}{c^2  a}  \, \frac{f_{k}(r)f_{k}(r_{s}-r)}{f_{k}(r_{s})} \,.
\end{equation}
The functions $a(t)$ and $t(r)$ are the scale factor and geodesic time for the background FLRW model, $r_{s}=r(z_{s})$ is the comoving position of the source at redshift $z_{s}$ and $f_{k}(r)=\sin(r\sqrt{k})/\sqrt{k},\; r,\;\sinh(r\sqrt{-k})/\sqrt{-k}$ depending if the curvature $k>,=,<0$, respectively.

Because in Eq.~\eqref{eq:kappa1} the contributions to the total convergence are combined additively, it is useful to decompose the density field into a sum of Fourier modes:
\begin{equation} \label{dek}
\delta_{M}(\bm{r},t) = \int_0^{\infty} {d^{3} \bm{k} \over (2 \pi)^{3}} \, e^{i \bm{k} \cdot \bm{r}} \, \delta_{M}(\bm{k},t) \,.
\end{equation}
Eq.~(\ref{dek}) can be used to separate the contributions to the convergence due to small-scale inhomogeneities 
from the ones due to large-scale inhomogeneities 
so that, analogously to Eq.~(\ref{pktot}), it is:%
\footnote{Note that the splitting is \textit{a priori} different as power spectrum and lensing are affected by non-linearities in a different way.}
\begin{equation} \label{kappatot}
\kappa(z) = \kappa_{L}(z) + \kappa_{H}(z) \,.
\end{equation}

\subsubsection{2-halo convergence PDF}

Given Eqs.~\eqref{PkL}, \eqref{eq:kappa1} and \eqref{dek}, it is straightforward to obtain the variance of the convergence PDF:
\begin{equation} \label{varkappal}
\sigma^{2}_{\kappa_{L}}=   \int_{0}^{r_{s}} \dd r \rho_{M0}^{2}G^{2}(r,r_{s}) \int_0^{k_{L}} {k \, \dd k \over 2 \pi}  P_{L}\big(k,z(r)\big) \,.
\end{equation}
The third moment (skewness) can be obtained via the bispectrum (three-point correlation function), which should also contribute to the variance (although the contribution is expected to be small at these linear scales).
Here, we take an alternative approach and model the full PDF relative to the 2-halo term as a shifted log-normal distribution with zero mean and the variance given by Eq.~\eqref{varkappal}, see \citet{Marra:2013roi} for details.
This is motivated by the fact that the lensing PDF inherits -- see Eq.~\eqref{eq:kappa1} -- the well-known log-normal character of the density contrast PDF \citep{Coles:1991if, Kayo:2001gu}.
As it can be seen by the example of Fig.~\ref{fig:12compa}, the final distribution is a minor modification to the corresponding Gaussian and captures the fact that, even in the linear regime, underdensities occupy more volume than overdensities, causing an intrinsic skewness in the distribution. See \citet{Barthelemy:2023mer} for a model from first-principles that captures the skewness of the PDF in the quasi-linear regime.

In Eq.~\eqref{varkappal}, the cut-off scale $k_L$ separates the 2-halo contribution, characteristic of larger scales, from the 1-halo term, relevant at smaller scales. The precise determination of $k_L$ is theoretically challenging, especially because, at a given redshift, the dependence on the halo masses is non-trivial. Consequently, we employ a pragmatic approach to establish this scale, which involves adjusting it so that the lensing PDF derived from \texttt{tgl} matches that obtained from \pinocchio${}_{\rm halos}$. A detailed description of the method involving \pinocchio${}_{\rm halos}$ is provided in Section \ref{sec:pin}. The latter's matter field representation closely matches indeed the one used by \texttt{tgl}, making it a suitable reference for this calibration. We determined that $k_L = \exp (3.9-4.6 z) \text{ Mpc}^{-1}$ achieves a good agreement, showing that $k_L$ increases at lower redshifts. This aligns with the fact that the halo model does not represent low-density filamentary structures, which become relatively more important at lower redshifts. Addressing this discrepancy necessitates the inclusion of higher mode wavenumbers in calculating the 2-halo variance, thereby compensating for the model's lack of power in these regions.

\subsubsection{1-halo convergence PDF}

The contribution to the convergence from the 1-halo term could be similarly computed via the halo power spectrum:
\begin{equation} \label{varkappah}
\sigma^{2}_{\kappa_{H}}=   \int_{0}^{r_{s}} \dd r \rho_{M0}^{2}G^{2}(r,r_{s}) \int_{k_H}^{k_{\rm max}} {k \, dk \over 2 \pi}  P_{H}\big(k,z(r)\big) \,,
\end{equation}
where $k_{\rm max}$ is the cut-off to account for the finite angular resolution. 
The other cut-off scale $k_H$ is instead relative to the size of the halos: for $k\ll k_H$, halos contribute to Eq.~\eqref{varkappah} only via shot noise.

Thanks to the sGL method, we can improve upon the previous result and obtain an estimate equivalent to including the information from all the $n$-point correlation functions.
As explained in \citet{Kainulainen2011a}, the halo contribution to Eq.~\eqref{eq:kappa1} can be rewritten as:
\begin{align} \label{eq:kH}
\kappa_H = \sum_{i,u} \,\kappa_{1iu}(z_s)  \Big(k_{iu}-\Delta N_{iu} \Big) \,,
\end{align}
where $\kappa_{1iu}$ is the convergence due to one object in the bin $i u$:
\begin{equation} \label{kappa1}
\kappa_{1iu}(z_s)\equiv  G_i(r_s) \, \Sigma_{iu}(t_i) \,,
\end{equation}
where $\Sigma_{iu}$ is the surface mass density of the halo in the bin $iu$.
Here, the index $i$ labels the lens planes, while the index $u$ labels the parameters that define the surface density; for halos these are the impact parameter and the halo mass. 
Now, as far as the 1-halo term is concerned, the integers $k_{iu}$ are distributed as Poisson random variables:
\begin{equation}
P_{k_{iu}} = \frac{(\Delta N_{iu})^{k_{iu}}}{k_{iu}!} \, e^{-\Delta N_{iu}} ,
\label{eq:poisson-i}
\end{equation}
where the parameter $\Delta N_{iu}$ is the expected number of objects in the bin volume $\Delta V_{iu}$:
\begin{equation} 
\Delta N_{iu} = \Delta n_{iu} \, \Delta V_{iu} = \Delta n_{iu} \, \Delta r_i\, \Delta A_{iu} \, ,
\label{deltan}
\end{equation}
where $\Delta n_{iu}$ is the comoving density of objects corresponding to the parameters in the bin $iu$, $\Delta r_i$ is the bin in the geodesic path, and $\Delta A_{iu}$ is the corresponding cross-sectional area of the halos in co-moving units.

From Eq.~\eqref{eq:kH}, using the properties of the Poissonian distribution, one can see that $\langle \kappa_H \rangle = 0 $ and \citep{Kainulainen2011a,Kainulainen2011b}:
\begin{align}
    \sigma^{2}_{\kappa_H}&=   \sum_{i,u} \,\kappa_{1iu}^{2} \,  \Delta N_{iu} \,,
    \label{varkappa} \\
    \mu_{3, \kappa_H}&=  \sum_{i,u} \,\kappa_{1iu}^{3} \,  \Delta N_{iu} \,,
    \label{skewkappa}
\end{align}
from which, going back to integral form, one obtains:
\begin{align}
    \sigma^{2}_{\kappa_H}&=   {1 \over N_{O}} \int_{0}^{r_{s}} \dd r \, G^{2}(r,r_{s}) \int_{M_{\rm min}}^{\infty} \dd n (M,z(r))  \nonumber \\
    & \times \int_{0}^{R(M,z(r))}\dd A(b,M)   \, \Sigma^{2}(b,M,z(r)))  \,,  \\
    \mu_{3, \kappa_H}&=   {1 \over N_{O}} \int_{0}^{r_{s}} \dd r \, G^{2}(r,r_{s}) \int_{M_{\rm min}}^{\infty} \dd n (M,z(r))  \nonumber \\
    & \times \int_{0}^{R(M,z(r))}\dd A(b,M)  \, \Sigma^{3}\big(b,M,z(r)\big)  \,. 
\end{align}
The integral limits for the last two integrals are implicitly defined, $\d A(b) \equiv 2 \pi b \, \d b$ and $\Sigma$ is the halo surface density:
\begin{equation}
\Sigma(b,M,z) = a^{3} \int_{b}^{R} \frac{2 \,   r \, {\rm d}r}{\sqrt{r^2-b^2}} \, \rho(r,M,z) \,.
\end{equation}
Again, we want to point out that these expressions use the halo profiles in real space and not in Fourier space, thus including higher-order correlation terms beyond the power spectrum.

Obtaining the moments beyond $\mu_3$ analytically is more challenging as there is no closed form. It is easier to estimate them by drawing a set of Poissonian numbers $k_{iu}$, from which the full convergence PDF and its moments can be easily estimated.

\subsubsection{Full convergence PDF}

The full convergence PDF is finally obtained by convolving the 1- and 2-halo PDFs.
The former is obtained via a histogram of a (Poissonian) realization; the latter is based on the log-normal template previously discussed. The final PDF is in the image plane.

The 1-halo contribution is based on binning the comoving space around the photon geodesic. We adopt a fine enough binning such that the numerical results converge. Furthermore, we smooth the NFW density profile via a Gaussian kernel in order to have a fixed angular resolution. The PDF will then be relative to this corresponding smoothing angular resolution, as is the case for the other methods adopted in this study.
Finally, the modeling of the halos follows the prescription adopted in \pinocchio; that is, we adopt the mass function by \citet[][Eq.~12, FOF]{2013MNRAS.433.1230W} and the concentration parameter model by \citet[][$200_c$]{2013ApJ...766...32B}. Furthermore, the smallest halo we model ($M_{\rm min}$ in the previous equations) has a mass that corresponds to the minimum halo mass adopted in the \pinocchio\  simulations, that is, $2.6 \times 10^{9} h^{-1} \Msun$ (10 dark matter particles, see Table~\ref{tab:sims}).

\subsection{Light-cone construction and ray tracing}

The past light cone (PLC) lens maps for the \pinocchio\ and $N$-body simulations were constructed using the \texttt{SLICER}\footnote{\url{https://github.com/TiagoBsCastro/SLICER}} code \citep{tiago_castro_2024_11048430}, introduced in~\citet{Castro2018}.\footnote{\texttt{SLICER} is an improved and fully rewritten version of \texttt{MapSim} \citep{Giocoli:2015tka}.}
This code distributes particles within the light cone based on the simulation snapshots. The process begins with \texttt{SLICER} configuring the field of view and setting the maximum source redshift, as specified in the input file. Subsequently, the code determines the required number of lens planes to ensure a continuous construction of the light cone.
Each simulation snapshot is then processed sequentially. The code reads the files, selecting particles within the predetermined field of view. Prior to projecting these particles onto the lens plane, each snapshot undergoes a randomization process. This step involves reflecting and translating the center of the particle distribution, in accordance with the periodic boundary conditions. Additionally, a specific face of the simulation cube is chosen to align with the line of sight.
The final stage involves constructing the lens planes. This is achieved by mapping the positions of particles to the nearest preselected lens plane. The angular positions of the particles are maintained during this process. The surface density is then pixelated using the triangular cloud method, as described in \citet{Hockney1998,Bartelmann2003}. The grid pixels are carefully selected to maintain a consistent angular size across all planes.
To prevent repetition of cosmic structures in the field of view (FoV) and to ensure that the map sizes are equivalent to the simulation box size (set at $150 \textrm{ Mpc}/h$), all maps are constructed with a uniform FoV value of $1^\circ$, see Table~\ref{tab:sims}. Consequently, using \texttt{SLICER}, we successfully generated 20 light cones up to $z_{s} = 5$. These cones comprise 118 lens maps with varying resolutions: $2048^2$, $1024^2$, $512^2$, and $256^2$ pixels. These resolutions correspond to angular resolutions of $1.76$, $3.52$, $7.03$, and $14.06 \textrm{ arcsec}$, respectively.

\begin{table}
\caption{Basic properties from the simulations used in this work. From left to right: the size of the box, gravitational softening, number of DM particles and their masses, the number of lens planes built up to $z=5$, the field of view of the constructed past light-cone, and the number of PLC realizations (different random seeds).}
\label{tab:sims}
\centering
\renewcommand{\arraystretch}{1.3}
\setlength{\tabcolsep}{4pt}
\begin{tabular}{ccccccc}
    \hline\hline
    $L_\textrm{box}$ & $\epsilon_{\rm{soften.}}$ & $N_\textrm{part.}$ & $m_\textrm{DM}$ &  $N_{\rm planes}$ & FoV & PLCs\\
    {\rm (Mpc/h)}& (kpc/h)& & $(\Msun /h)$ & ($z\le5$) & (deg.) & \\
    \hline
     $150$ & $1.4$  & $1024^3$ & $2.6\times10^8$ & $118$ & $1.0$ & 20 \\
    \hline
\end{tabular}
\medskip
\end{table}

After creating the lens planes, the lensing quantities (convergence, $\kappa$, shear, $\gamma$, and magnification, $\mu$) are computed using a ray-tracing technique that integrates the non-linear matter distribution along the unperturbed trajectory from the source plane to the observer. \citet{Castro2018} showed that, under the Born approximation, the 1-point lensing statistics deviates by less than 10\% up to $z=5$ compared to the full solution within the 68\% highest density interval. However, the tails of the distribution may be significantly affected. We independently validate the degree of validity of the Born approximation by comparing the results when using \glamer\ deflection solver~\citep{Metcalf:2013kya,Petkova:2013yea}. Nevertheless, since our focus is on the relative differences between simulation methods, and we use the same ray-tracing algorithm for all simulations, the relative differences remain basically unaffected.
In the thin lens approximation, i.e., disregarding the size of each lens, the convergence for a single lens plane is
\begin{equation}
    \kappa\left(\bm{\theta}\right) = \frac{\Sigma\left(\bm{\theta}\right)}{\Sigma_{\rm{crit}}},
 \label{eq:kappa}
\end{equation}
where $\bm{\theta}$ is the observed angular position, $\Sigma\left(\bm{\theta}\right)$ is the surface mass density and $\Sigma_{\rm{crit}}$ is the critical surface density
\begin{equation}
    \Sigma_{\rm{crit}} = \frac{c^{2}D_{l}}{4\pi G D_{s}D_{ls}}.
\end{equation}
In the above, \textit{c} is the speed of light, \textit{G} is the gravitational constant, $D_{l}$, $D_{s}$ and $D_{ls}$ are the angular diameter distances between the observer-lens, observer-source, and lens-source, respectively.
Thus, we use equation \eqref{eq:kappa} to obtain the differential convergence maps, which are summed up to the source redshift.

The shear maps were computed using the Shear Map Reconstruction (SMR) \texttt{PYTHON} script distributed with \texttt{SLICER}. The shear components $\gamma_{1}, \gamma_{2}$ are written as a function of the deflection potential $\Psi$:
\begin{align}
\gamma_{1}\left(\bm{\theta}\right) &=  \frac{1}{2}\left(\Psi_{\theta_{1}\theta_{1}}-\Psi_{\theta_{2}\theta_{2}}\right),\\
\gamma_{2}\left(\bm{\theta}\right) &= \Psi_{\theta_{1}\theta_{2}} = \Psi_{\theta_{2}\theta_{1}},
\end{align}
where 
\begin{equation}
    \Psi\left(\bm{\theta}\right) = \frac{1}{\pi}\int \kappa\left(\bm{\theta'}\right)\ln\left|\bm{\theta}-\bm{\theta'}\right|\d^{2}\theta'\,.
\end{equation}
SMR uses FFT to reconstruct the lensing potential from the convergence maps and fourth-order finite differences to calculate its shear components. While tacitly assuming periodic boundary conditions when using FFT to reconstruct the shear maps is not formally correct, we have assessed its impact on the lensing PDF and found it to be completely subdominant compared to the differences between the simulation methods.

Finally, the magnification $\mu$ is computed using the relation
\begin{equation}
    \mu \equiv \frac{1}{\left(1-\kappa\right)^{2}-\gamma^{2}},
\end{equation}
where $\gamma = \sqrt{\gamma_{1}^{2} + \gamma_{2}^{2}}$ is the shear modulus. In the weak lensing regime, the above relation is reduced to
\begin{equation}
    \mu\approx 1+2\kappa+\gamma^{2}+3\kappa^{2}.
\end{equation}

We computed the angular power spectra $P_{\kappa}$ and $P_{\gamma}$ for all the convergence and shear maps, respectively. In linear theory, it is $P_{\kappa} = P_{\gamma}$, and for a source at $z_{s}$ it is written as an integral of the density contrast along the unperturbed path of light, using the Limber approximation \citep{Limber1953, Munshi2008, Bartelmann2001}:
\begin{equation}
    P_{\kappa}\left(l,z_{s}\right) \!\!=\!\!  \frac{9H_{0}^{4}\Omega_{\rm m0}^{2}}{4c^{4}}  \!\!\int_{0}^{\chi_s} \!\!\!\!\! \dd\chi' \! \left(\frac{\chi_s-\chi'}{\chi_s \, a(\chi')}\right)^{2}\!\!\! P_{m}\left(\frac{l}{\chi'},\chi'\right),
\end{equation}
where $\chi(z)$ is the comoving distance at $z$, $\chi_s = \chi(z_{s})$, and $P_{\rm m}$ is the linear matter power-spectrum.

\subsection{\pinocchio}\label{sec:pin}

\pinocchio\  generates simulated halo catalogs using the ellipsoidal collapse to compute the collapse time, defined as the moment of orbit crossing, and Lagrangian perturbation theory (LPT) to displace the formed halos and uncollapsed particles. Briefly, \pinocchio\  works as follows: a linear field is smoothed using a Gaussian filter at various scales, the collapse time is calculated for each particle (grid point) with the ellipsoidal collapse; these are then sorted by collapse time in chronological order and displaced according to the LPT theory; later, the code groups the collapsed particles into halos and filaments in order to properly reproduce the hierarchical formation of the structures; and finally the LPT theory is used to move the halos to their final position and thus describe their evolution. As the particles that are members of a halo have passed through orbit crossing, their displacements are extrapolations of the LPT displacement outside its validity; therefore, \pinocchio\  cannot provide reliable profiles needed for accurate lensing statistics.
The particles belonging to halos are then redistributed assuming a NFW profile and the mass-concentration relation given by~\citet[][$200_c$]{2013ApJ...766...32B}.

We use \pinocchio\  configured in the third order of the LPT theory to build a set of simulations defined according to Table \ref{tab:sims}. In an effort to better compare \pinocchio\ and sGL results, we consider two \pinocchio\  configurations (see Table~\ref{tab:summary}):
i) the label \pinocchio\  represents the complete results (halos+filaments+uncollapsed particles that do not belong to any structure) of the standard \pinocchio\ code, and
ii) the label $\pinocchio_{\textrm{halos}}$ considers only the halos and randomizes the rest.

As a note, the construction of a lightcone from snapshots is an inefficient use of \pinocchio, because the code itself outputs halos in a lightcone, and the position of particles outside halos can be easily recovered at any time using LPT. However, in this study, we prefer to apply to the \pinocchio\ runs the same analysis applied to the $N$-body simulations.

\subsection{$N$-body simulations}

The $N$-body simulations used in this work were performed using the GADGET-2 code \citep{Springel2005}, following the configurations described in Table~\ref{tab:sims}. The GADGET-2 code uses the parallelized Tree–Particle Mesh~\citep[TreePM, see][]{Xu:1994fk} algorithm to calculate the gravitational interactions between the simulation particles. In this algorithm, these interactions are separated into a long-range term, calculated using the Particle-Mesh methods, and a short-range term for interactions of close neighbors that are described via the Tree algorithm \citep{Barnes1986nb}.

The simulations contain only collisionless matter accounting for the amount of cold dark matter and baryons, excluding any effect of baryonic feedback. The dark matter particles were evolved from $z=99$ to $z=0$ in a comoving box with a side equal to 150 Mpc/h and mass resolution of $2.6\times 10^{8}\,\Msun/h$ (see Table \ref{tab:sims}). The choice for this box size is justified by the fact that it has already been shown in~\citet{Castro2018} that a box of 128 Mpc/$h$ with a mass resolution of $10^{8}\,\Msun/h$ is enough to obtain convergent results for the lensing PDF.

\subsection{Hydrodynamical simulations}\label{sec:hydro}

The effect of baryons on lensing probability density functions and the convergence power spectrum \(P_\kappa(\ell)\) was thoroughly examined by \citet{Castro2018} using hydrodynamic simulations. This study specifically analyzed the multipole range of \(P_\kappa(\ell)\) and the value ranges for \(\kappa\), \(\gamma\), and \(\mu\) across various redshifts, identifying thresholds below which the baryonic contributions are minimal. Given the challenges associated with extracting cosmological information in scenarios where baryonic effects—such as AGN and stellar feedback, star formation, magnetic fields, and thermal conduction—are significant, our analysis adopts a simplified approach. We thus avoid the complexities of expensive hydrodynamical simulations by focusing on the dark matter-dominated regime, ensuring that baryonic effects remain below the 50\% threshold.
Table~\ref{tab:DM-ranges} displays the ranges within which the PDFs from the Hydro and DM versions of Box 3 (hr) of the \textit{Magneticum} simulation suite differ by less than 50\%, given a resolution of 1.76 arcsec.
Although more stringent thresholds could be employed, as discussed in the next section, these values effectively limit our discussion to the bulk of the distribution, which is well described by a DM only model.

\begin{table}
\centering
\caption{Ranges where the PDFs from the Hydro and DM versions of Magneticum Box 3 hr differ by less than 50\% (1.76 arcsec).}
\label{tab:DM-ranges}
\setlength\tabcolsep{6pt}
\renewcommand{\arraystretch}{1.0}
\begin{tabular}{ccccc}
    \hline
    \hline
    \textbf{Statistics} & \textbf{Redshift} & \textbf{Range} & $\mathcal{L}_1$ [\%] \\
    \hline
    Convergence & 1 & [$-\infty$, 0.38] & 1.3\\ 
    Convergence & 2 & $[-0.12, 0.50]$ & 1.9\\ 
    Convergence & 3 & $[-0.17, 0.50]$ & 2.5\\ 
    Convergence & 5 & $[-0.25, 0.49]$ & 2.9\\ 
    \hline
    Shear & 1 & [$1.0 \times 10^{-4}$, $3.5 \times 10^{-2}$] & 2.1\\ 
    Shear & 2 & [$1.0 \times 10^{-4}$, $6.0 \times 10^{-2}$] & 3.0\\ 
    Shear & 3 & [$1.0 \times 10^{-4}$, $8.0 \times 10^{-2}$] & 3.7\\ 
    Shear & 5 & [$1.0 \times 10^{-4}$, $1.0 \times 10^{-1}$] & 4.1\\ 
    \hline
    Magnification & 1 & $[0.93, 2.3]$ & 1.3\\ 
    Magnification & 2 & $[0.88, 2.3]$ & 1.9\\ 
    Magnification & 3 & $[0.83, 2.4]$ & 2.4\\ 
    Magnification & 5 & $[0.72, 2.9]$ & 2.8\\ 
    \hline
    Magnitude & 1 & $[-1.17, 0.11]$ & 1.3\\ 
    Magnitude & 2 & $[-1.34, 0.24]$ & 2.0\\ 
    Magnitude & 3 & $[-1.52, 0.32]$ & 2.5\\ 
    Magnitude & 5 & $[-1.52, 0.46]$ & 2.9\\     
    \hline
\end{tabular}
\bigskip
\end{table}

\subsection{The $\mathcal{L}_1$ norm}\label{sec:L1}

The PDFs of Hydro and DM simulations differ more significantly on the tails; therefore, a local discrepancy of 50\% usually corresponds to a much smaller global change of statistics. It is, hence, interesting to compute the weighted relative difference between a surrogate and reference PDFs:
\begin{equation}
    \mathcal{L}_1 (a, b) \equiv \frac{1}{2}\int_a^b{\frac{\left|\frac{\dd P}{\dd x}_{\rm model} - \frac{\dd P}{\dd x}_{\rm surrogate}\right|}{\frac{\dd P}{\dd x}_{\rm model}}\,\frac{\dd P}{\dd x}_{\rm model} \dd x}\,,
\end{equation}
that tautologically reduces to
\begin{equation}
    \mathcal{L}_1(a, b) = \frac{1}{2}\int_a^b{\left|\frac{\dd P}{\dd x}_{\rm model} - \frac{\dd P}{\dd x}_{\rm surrogate}\right| \dd x}\,,
    \label{eq:l1}
\end{equation}
which gives the cumulative absolute difference between the distributions.
We refer to this statistic as the $\mathcal{L}_1$ norm, in analogy to the metric with the same name for distance between two vectors.

In particular, $\mathcal{L}_1(-\infty,\infty)$ is a simple scalar that quantifies the global concordance of two PDFs. We will refer to $\mathcal{L}_1(-\infty,\infty)$ as simply $\mathcal{L}_1$ in what follows. In the limit in which both PDFs have zero overlap, we get $\mathcal{L}_1 = 1 = 100\%$, which can be interpreted as 100\% of the observed histogram bins of the surrogate PDF will be misplaced with respect to the model. Hence the prefactor of 1/2 in Eq.~\eqref{eq:l1}.
The last column of Table~\ref{tab:DM-ranges} shows $\mathcal{L}_1(a,b)$, relative to the range given in the third column, between the surrogate (DM) and reference (Hydro) PDFs.
We see that a local discrepancy of $50\%$ in the PDFs corresponds to a weighted relative difference of 2\% for magnification.


\begin{figure}
    \centering
    \includegraphics[width=.96\columnwidth]{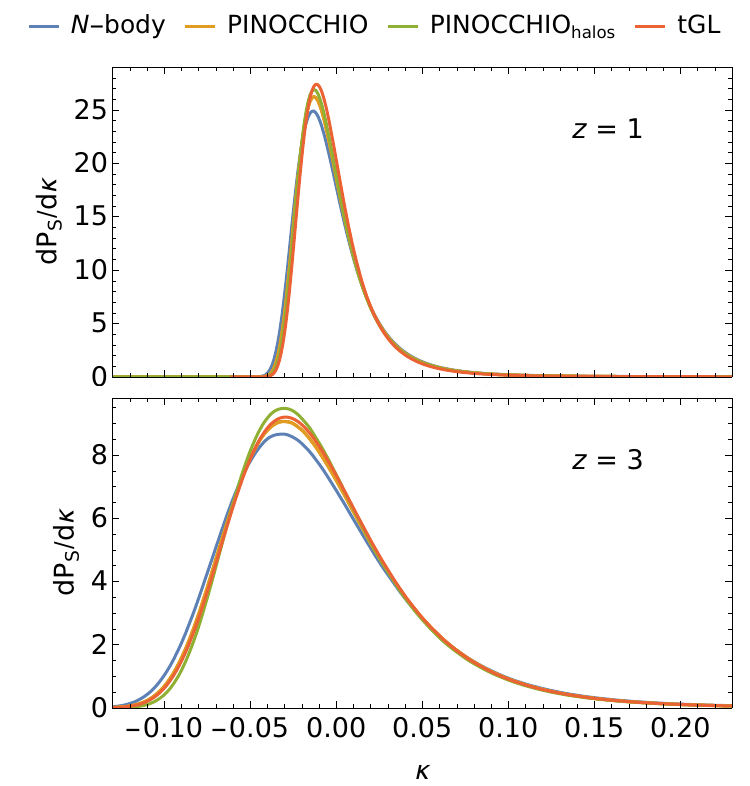}
    \caption{Convergence PDFs for the different lensing methods considered in this study, for an angular resolution of 1.76 arcsec. The Poisson errors in the PDFs were not shown because their value is smaller than the scale of the figure.}
    \label{fig:PDFs-linear-scale}
    \medskip
\end{figure}

\begin{figure*}
\includegraphics[width=\linewidth,trim=0 38 0 0 ,clip] {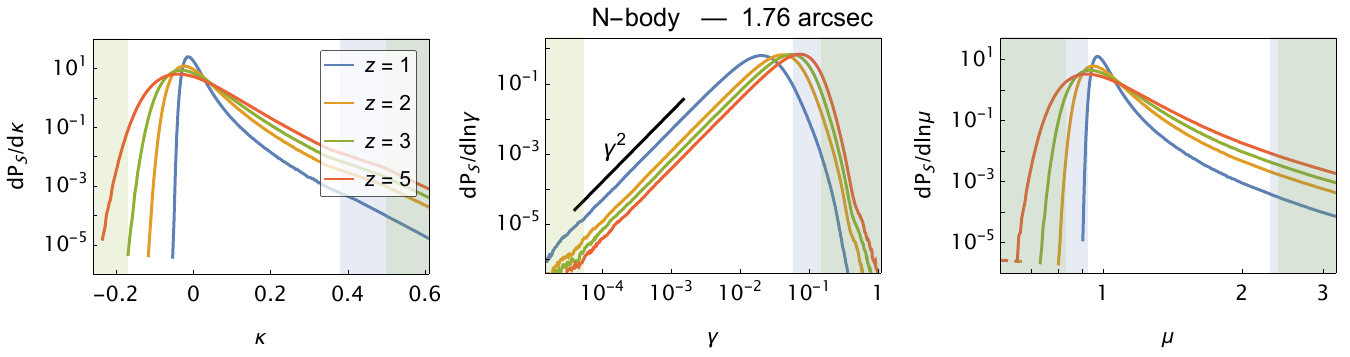}
\includegraphics[width=\linewidth] {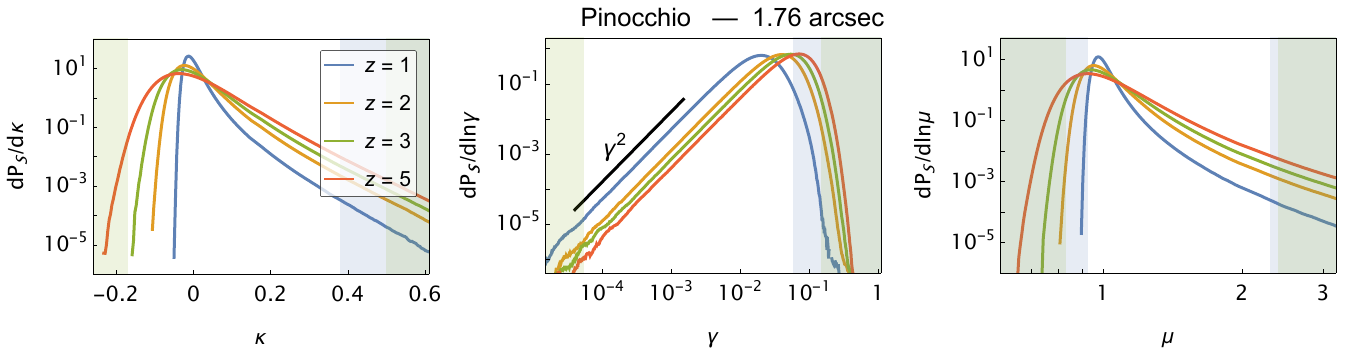}
\caption{Comparison of $\kappa$, $\gamma$ and $\mu$ PDFs (in the source plane) for different redshifts for the $N$-body and \pinocchio\ case for the angular resolution of 1.76 arcsec. $\pinocchio_{\textrm{halos}}$ (not shown here for compactness) exhibits similar behavior to $\pinocchio$. We use semi-log scale for $\kappa$ and log-log for $\gamma$ and $\mu$ for better clarity.
The shaded regions mark where baryonic corrections exceed 50\% for $z=1$ (blue) or for $z=3$ (green).    As can be seen, the two methods agree reasonably well in the range where baryonic effects are under control. See also Fig.~\ref{fig:residuals}.
\label{fig:pdf-comparison-zs}}
\medskip
\end{figure*}

\bigskip
\section{Results}\label{sec:test}

We now present our results for the PDFs of convergence, shear, magnification, and magnitude. In the ray-tracing method employed, rays are calculated from the source to the observer, resulting in statistics computed in the image plane. In most cases, we subsequently translate these PDFs to the source plane, which is relevant to observations. Hereafter, the PDFs in the image and source planes are denoted by the subscripts \textit{I} and \textit{S}, respectively. For the extraction of these PDFs, we construct histograms of the lensing quantities using different binning schemes suited to their respective distributions. The convergence is binned linearly over the fixed range [$-0.5 , 3.0$] using one thousand bins. For the shear and magnification, logarithmic binning is applied to account for their wide dynamic ranges and to emphasize the behavior in the tails, with respective ranges of [$1\times10^{-6},2\times10^0$] and [$1\times10^{-1},1\times10^2$], while maintaining the same number of bins as for the convergence. We computed all PDFs for redshifts $z = 1,\,2,\,3,$ and 5; however, to avoid overcrowding in most plots, only the results for $z=1$ and $z=3$ are shown.

As detailed in \cite{Castro2018}, the lensing PDFs intrinsically depend on the angular resolution. This dependency is not directly related to the point spread function of any particular telescope, but rather to the angular scale used for computing averages. For example, for a fixed lens and source position, the average magnification of galaxies with diameters of 1 and 5 arcsec may differ. As discussed in Section~\ref{sec:hydro}, our simulations yield reliable results for a minimal angular resolution of 1.76 arcsec, which we adopt as our fiducial resolution for most discussions. However, results for lower resolutions are presented in Appendix~\ref{app:resolutions}.

\subsection{Lensing PDFs}\label{sec:pdfsec}


In Fig.~\ref{fig:PDFs-linear-scale}, we display the $\kappa$ PDFs derived using the three different methods considered in this study: full $N$-body simulations, \pinocchio\ (including the `halos' variant), and the stochastic \texttt{turboGL} code, as described in Section \ref{sec:lenssims} and summarized in Table~\ref{tab:summary}. These are depicted for $z=1$ and $z=3$ at our fiducial angular resolution of 1.76 arcsec. The methods generally exhibit good agreement, particularly in the peak position. However, the $N$-body simulations show a slightly larger variance, as indicated by the lower PDF peak and normalization condition. Since these plots probe weak convergences and, therefore, mildly nonlinear structures, the slight discrepancies should be attributed to a lack of modeling precision at the scale of filaments, which will be discussed subsequently.

However, differences in the tails of the distributions are obscured by the linear scale used in Fig.~\ref{fig:PDFs-linear-scale}. Therefore, in Fig.~\ref{fig:pdf-comparison-zs}, we present a general comparison between the results from $N$-body and \pinocchio\ for $\kappa$, $\gamma$, and $\mu$ on a logarithmic scale, depicting four different redshifts for each method. The \texttt{turboGL} code is excluded from this comparison as it does not compute the shear PDFs so that $\mu$ can be inferred only in the weak-lensing regime. As previously discussed, our analysis focuses on the region where baryonic effects are maintained below the 50\% level. The regimes where this is not the case at $z=1$ and $z=3$ are highlighted with a blue and green shade, respectively. It is evident that as the source distance increases, the convergence, shear, and magnification PDFs become wider (indicating higher variance), and the peaks of their distributions shift (indicating skewness). We will further explore these trends for convergence in Section~\ref{dissection}, noting the general trend that lensing intensifies at higher redshifts due to the increased number of lens encounters. We find that both \pinocchio\ and $N$-body simulations are consistent with the theoretical model of randomly distributed point-mass lenses~\citep{Schneider1992}, which predicts that when $\gamma < 0.01$, the PDF is proportional to $\gamma^2$, and for $\mu \gg 1$, the PDF is proportional to $\mu^{-1}$. However, while the asymptotic behavior prediction of the randomly distributed points model for the $\gamma-$PDF  happens within the range where the baryonic effects are subdominant, the prediction for the $\mu-$PDF occurs in the regime deeply dominated by baryonic physics and is not shown in Fig.~\ref{fig:pdf-comparison-zs}. The behavior of the magnification PDF at this regime will be discussed in Sect.~\ref{app:stronglens}.
We observe a good agreement between \pinocchio\ and $N$-body for the shear, with differences becoming apparent at $\gamma \gtrsim 0.1$, a region dominated by baryonic effects. An explanation for this disagreement will be proposed in Section~\ref{sec:maps}.

\begin{figure}
$\,$\includegraphics[width=.97\linewidth]{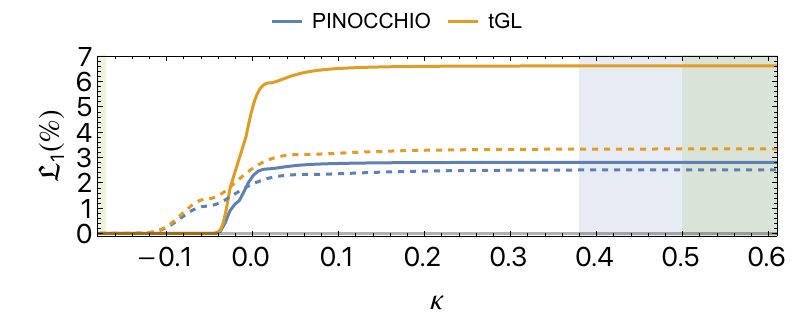}
\includegraphics[width=\linewidth, trim=12 0 0 0 ,clip]{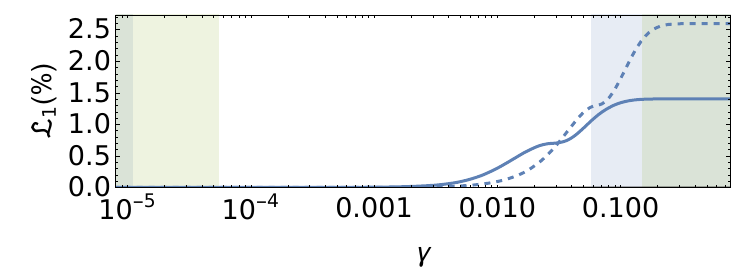}
\includegraphics[width=\linewidth, trim=3 0 0 0 ,clip]{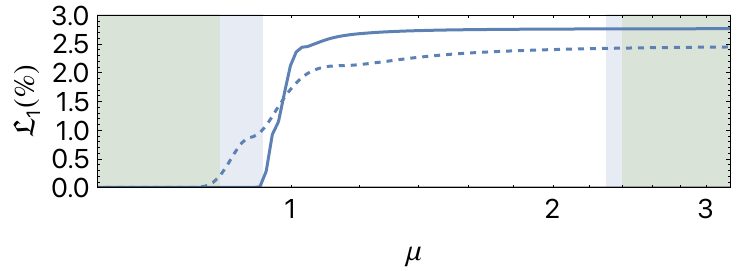}
\caption{The $\mathcal{L}_1(-\infty,x)$ norm, where $x = \kappa,\,\gamma$ or $\mu$, computed between approximate methods and the full $N$-body results in the source plane, for $z=1$ (solid lines) and $z=3$ (dashed lines). The blue (green) shaded regions represent the interval in which the baryonic correction is over 50\% for $z=1$ ($z=3$).  
\label{fig:residuals}}
\medskip
\end{figure}

Figure~\ref{fig:residuals} uses $\mathcal{L}_1(-\infty,x)$ to illustrate the differences between the PDFs from \pinocchio\ and \texttt{turboGL} compared to the $N$-body results.  
Regarding convergence, \pinocchio\ exhibits a small excess of probability density at the peak of the distribution, leading to a 3\% relative error. \texttt{turboGL} shows similar agreement at $z=3$, which becomes worse at $z=1$.  
Regarding magnification, \pinocchio\ again shows a 3\% error. 
The discrepancy on the shear PDF is around 2\%, and roughly half of that if we limit $\gamma \lesssim 0.1$.

We show in Table~\ref{tab:L1norm} a comprehensive comparison of the various approximations. The first two lines summarize the results of Figure~\ref{fig:residuals}. The third line shows that the ${\cal{L}}_1$ norm is not sensitive to the tails of the distributions, where type 2 and 3 images are relevant (see next Section).
The fourth line shows that non-collapsed matter plays a smaller but non-negligible role in the lensing PDFs, especially at higher redshifts (see also Section~\ref{sec:maps}).
The fifth line shows that the use of the Born approximation leads to further small deviations for the bulk of the PDFs. These are smaller changes than those introduced by \pinocchio, but if one is interested in recovering the full accuracy of an expensive Hydro simulation, we find that it may be appropriate to adopt a full, no Born method.
Finally, the last line shows that, compared to the difference between $N$-body and \pinocchio, the difference between Hydro and $N$-body is similar for $\kappa$ and $\mu$, and three times as large for $\gamma$. 

The larger difference between hydro and $N$-body simulations for $\gamma$ compared to $\kappa$ and $\mu$ observed in Table~\ref{tab:L1norm} is due to the contributions of multiple images events (discussed in the next section). For the shear PDF, these contributions cluster over a small dynamic range $0.1\lesssim \gamma < 1$, while for the other variables, they are sparsed over a much more extensive range. Therefore, the crossing of the $50\%$ threshold is steeper for shear than for the other variables, causing a more significative impact in calculating $\mathcal{L}_1$.

\setlength\tabcolsep{4.5pt}
\renewcommand{\arraystretch}{1.5} 
\begin{table}
\caption{Values of the ${\cal{L}}_1$ norm for the different cases.}
\label{tab:L1norm}
\centering
\begin{tabular}{cc|ccc|ccc}
\hline
\hline
Reference & Surrogate & \multicolumn{3}{c|}{$z=1\,[\%]$}  & \multicolumn{3}{c}{$z=3\,[\%]$}\\
\cline{3-8}
       &  & $\kappa$ & $\gamma$ & $\mu$ & $\kappa$ & $\gamma$ & $\mu$  \\
\hline
$N$-body & \texttt{tgl}   & 6.6  &  -- & --  & 3.3 & -- & -- \\ 
$N$-body & \texttt{PINOCCHIO}  & 2.8 &  1.4 &  2.8 & 2.5  &   2.6 &   2.5 \\
\multirow{2}{*}{\vspace{+4pt}$N$-body} & \multirow{2}{*}{\vspace{+4pt}\texttt{PINOCCHIO}}  & \multirow{2}{*}{2.8} & \multirow{2}{*}{1.4} &   \multirow{2}{*}{2.8}  & \multirow{2}{*}{2.5} & \multirow{2}{*}{2.6} &   \multirow{2}{*}{2.5} \\
\multicolumn{2}{c|}{(only type I)} & & & & & & \\
\texttt{PINOCCHIO} & \texttt{PINOCCHIO}${}_{\rm halos}$ & 1.1 &  0.5 &  1.1 & 2.0  &  1.5 &  1.9 \\
no Born & Born & 1.3 & 1.2 & 1.3 & 1.0 & 1.6 & 1.1 \\
Hydro              & $N$-body      & 1.3 &  4.3 &  1.3 & 2.6 & 7.3 & 2.6 \\
\hline
\end{tabular}
\medskip
\end{table}

\subsection{Strong lensing PDFs}
\label{app:stronglens}

\begin{figure*}
    \centering
    \includegraphics[width=0.96\linewidth,trim=55 0 55 0 ,clip]{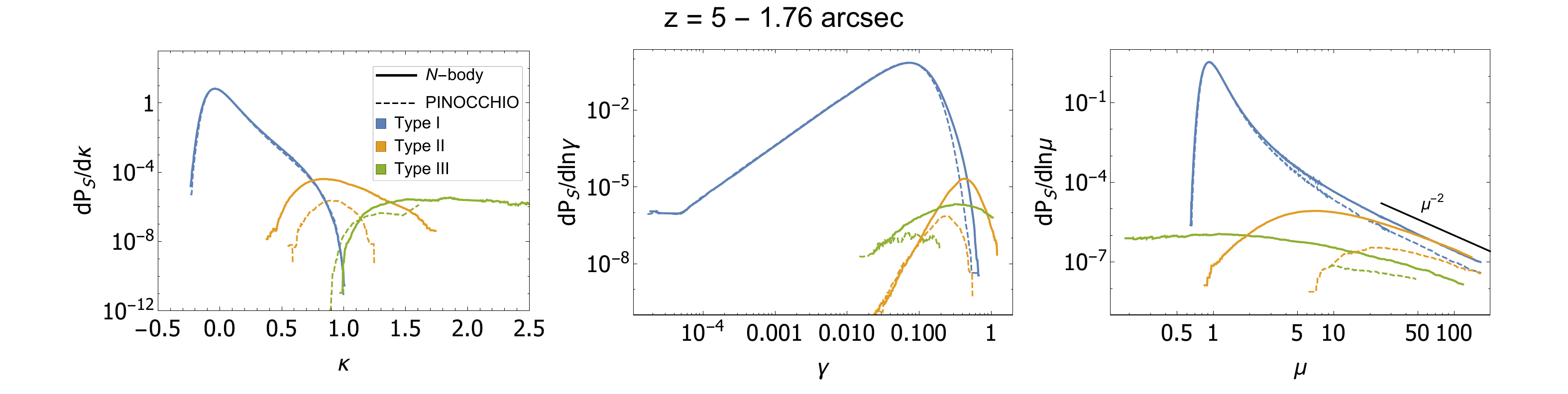}
    \caption{Convergence, shear, and magnification PDFs for different image types at $z = 5$ and an angular resolution of 1.76 arcsec.
    }
    \label{fig:pdf-typeI-III}
    \medskip
\end{figure*}

Here, we present the lensing PDFs in the strong-lensing regime, computed by differentiating contributions from various image types defined as follows:
\begin{align}
    \textrm{Type I: }& |1-\kappa|>\gamma\;\; \rm{and}\;\; \kappa<1 \,, \nonumber\\
    \textrm{Type II: }& |1-\kappa|<\gamma \,,\nonumber \\
    \textrm{Type III: }& |1-\kappa|>\gamma\;\; \rm{and}\;\; \kappa>1 \,.
\end{align}
Type II and III images indicate the strong-lensing regime, characterized by multiple images \citep{Schneider1992}, while Type I involves only image amplification.

Figure~\ref{fig:pdf-typeI-III} displays the PDFs of $\kappa$, $\gamma$, and $\mu$ for $z=5$, categorized into these three types.
We see that the $N$-body simulations show a higher probability of forming Types II and III images compared to structures generated by the \pinocchio\ code, highlighting the limitations of the method in accurately describing the strong-lensing regime (in the present implementation, the halo internal structure is defined by a single spherical NFW profile without substructure).
The absence of most Type II and III images in \pinocchio\ is also consistent with observations from the previous section that the \pinocchio\ PDFs disagree with the $N$-body results for $\gamma \gtrsim 0.1$. 
Furthermore, we find that the angular resolution does not significantly impact the lensing PDFs in the strong-lensing regime, and comparisons between the \pinocchio\ and \pinocchio$_{\rm halos}$ PDFs show nearly indistinguishable results, except the Type II $\kappa$ PDFs.

Additionally, we corroborate several findings from \cite{Castro2018}:
\begin{itemize}
    \item The magnification PDFs from the $N$-body simulations indicate that the observed plateau in the demagnification region results from Type II images.
    \item The emergence of secondary peaks in the high shear region is due to the transition from Type I to Types II and III.
    \item The plateau in the high-convergence tail of the convergence PDFs, starting at $\kappa \sim 1$, stems from the presence of Type III images.
\end{itemize}

\medskip
\subsection{Dissecting the PDF}\label{dissection}

\begin{figure}[t!]
    \centering
    \includegraphics[trim={0 0 0 0}, clip, width=.96\linewidth]{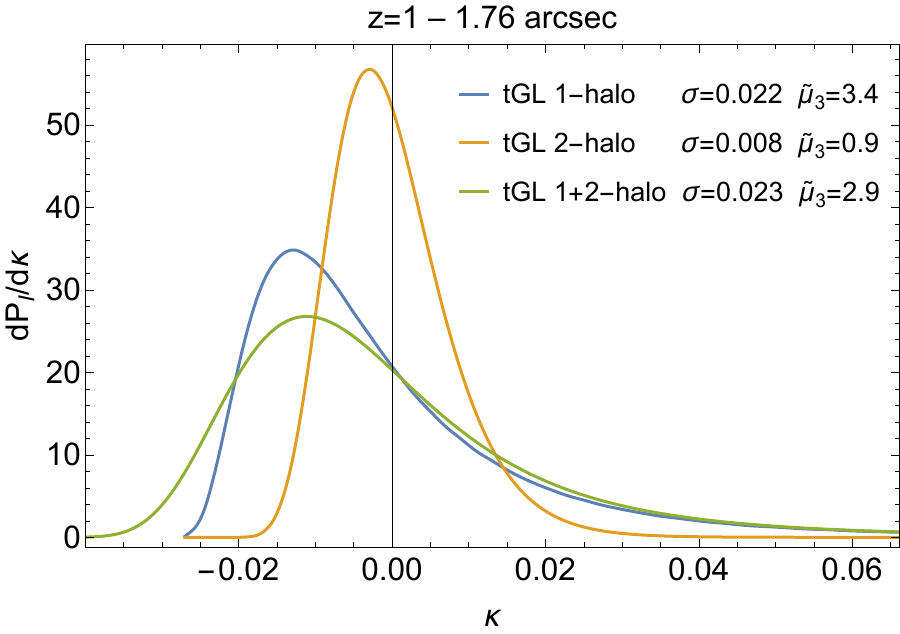}
    \caption{
    Image-plane convergence PDFs relative to the 1-halo and 2-halo contributions together with the full PDF obtained by convolving the previous two. One can see that the 1-halo contribution is more significant than the 2-halo one.
    \label{fig:12compa}}
    \medskip
\end{figure}

Figure~\ref{fig:12compa} shows the halo-model convergence PDF from \texttt{turboGL} together with the individual contributions from the 1- and 2-halo terms. We can see that the total lensing variance is dominated, as expected, by the 1-halo contribution and that the log-normal template features a mild skewness $\tilde\mu_3=\mu_3/\sigma^3$ (the third standardized moment). 

\begin{figure}[t!]
\centering
\includegraphics[trim={0 1.3cm 0 0}, clip, width=.96\linewidth]{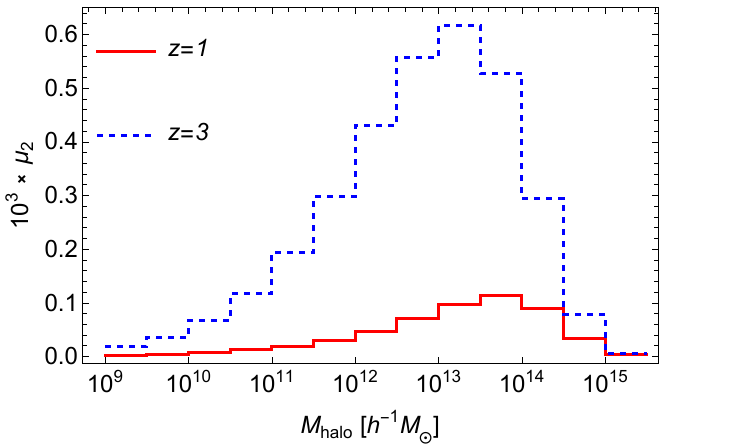}
\includegraphics[trim={0 0 0 0}, width=.96\linewidth]{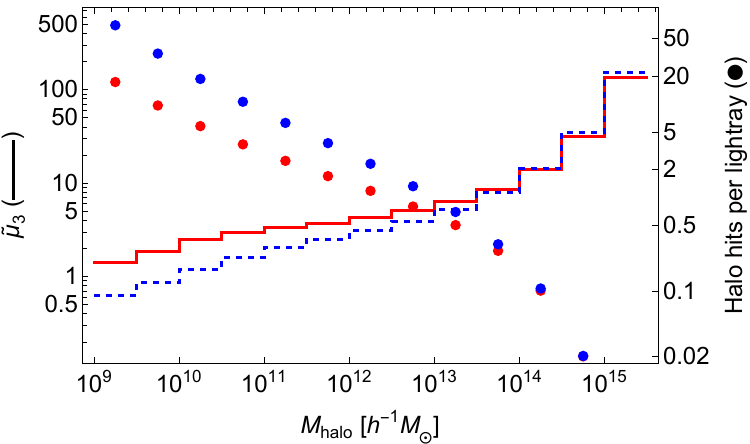}
\caption{
    Contribution to the variance $\mu_2$ (top) and skewness $\tilde\mu_3=\mu_3/\sigma^3$ (bottom) of lensing halos with a given mass for sources at $z=1$  and $z=3$ using Eqs.~(\ref{varkappa}-\ref{skewkappa}). Also shown (point markers) is the average number of lensing-halo hits per light ray. See Section~\ref{dissection}.
    \label{fig:diss-M}}
    \medskip
\end{figure}

\begin{figure}[t!]
\includegraphics[trim={0 1.1cm 0 0}, clip, width=.96\linewidth]{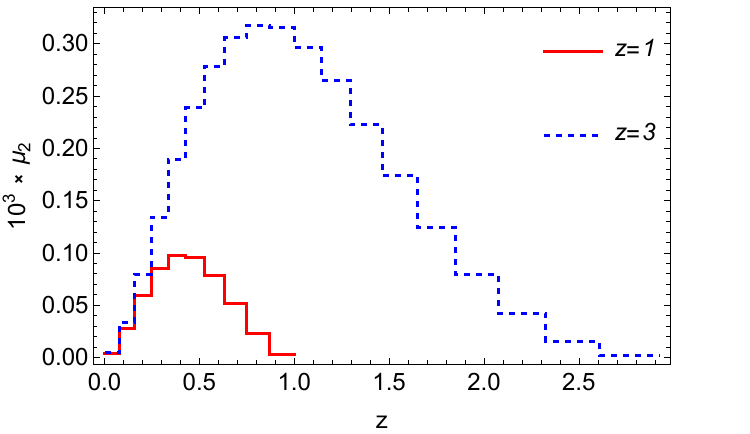}
\includegraphics[trim={0 0 0 0}, width=.96\linewidth]{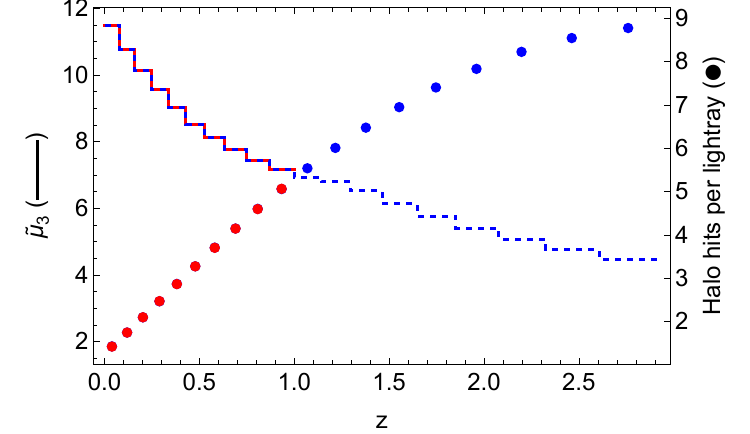}
\caption{
    Contribution to the variance (top) and skewness (bottom) of halos of a given lensing plane redshift for sources at $z=1$  and $z=3$  using Eqs.~(\ref{varkappa}-\ref{skewkappa}). Also shown (point markers) is the average number of lensing-halo hits per light ray.
    See Section~\ref{dissection}.
    \label{fig:diss-z}
}
\medskip
\end{figure}

Next, we show, breaking the sums of Eqs.~\eqref{varkappa} and \eqref{skewkappa} into individual elements, the contributions to the variance $\mu_2$ and skewness $\tilde\mu_3$ of halos with a given mass (Fig.~\ref{fig:diss-M}) and in a given lensing plane redshift (Fig.~\ref{fig:diss-z}), for sources at $z=1$  and $z=3$.
Here, we only consider the 1-halo term.
It is also shown the average number of lensing-halo hits per light ray.
In the case of Fig.~\ref{fig:diss-z}, a value of 10 in a given $\Delta z$, for instance, means that the light ray hit on average 10 halos in that redshift bin (bin widths are constant in comoving distance), and similarly for Fig.~\ref{fig:diss-M}.

From Fig.~\ref{fig:diss-M} (top panel) we can see that our simulations consider the relevant range of masses as the halos with mass $\approx 10^{13} h^{-1} M_\odot$ contribute to most of the variance. Halos with masses below $10^{9} h^{-1} M_\odot$ or above $10^{15} h^{-1} M_\odot$ have a negligible impact. This also means that the geometrical optics limit, which we assume here, should be valid for any source with wavelengths $\lesssim 10^{10}$m~\citep{Leung:2023lmq}, which covers not only supernovae but also future gravitational wave sources detected on the ground and on space with LISA. We also note that the lensing PDF for sources at $z=3$ has a variance that is about 5 times greater than the PDF for sources at $z=1$ -- the light ray encounters more lenses in the former case. Regarding skewness, we observe that more massive halos contribute significantly to the overall asymmetry of the PDF. This is due to less massive halos being more numerous, thereby distributing the same fraction of the Universe's mass across a greater number of objects. Consequently, a ray experiences a higher number of interactions (hits). According to the central limit theorem, this results in a PDF that more closely approximates a Gaussian distribution.
Finally, a source at $z=3$ is further away, and a larger number of halos are encountered so that the number of hits is larger and the skewness is lower.
Additionally, in this case, the lensing efficiency peaks at a higher redshift — $z \approx 1.5$ compared to $z \approx 0.5$ — making lensing more sensitive to structures that are more linear and on larger scales, thus closer to Gaussianity.
At the very high-mass end, the skewness for sources at $z=3$ is slighter higher because the extra number of hits with respect to $z=1$ does not compensate for the very skewed distributions of those very massive halos.

The top panel of Fig.~\ref{fig:diss-z} shows the contributions of the lens planes to the variance. The shape is modulated by the lensing efficiency of Eq.~\eqref{opw} and the growth of structures. The largest contributions come at about half way between source and observer. The bottom panel shows the contribution to the skewness.
The overall trend is that lens planes at lower redshifts contribute to a higher skewness. From Fig.~\ref{fig:diss-M}, we remember that smaller halos have a smaller number of hits at lower redshifts (and produce relatively more skewness), while this is the opposite for larger halos. The overall effect is dominated by the smaller halos, which can be attributed to the progressive nonlinearity in the Universe due to the growth of structures.
We also see that the contributions for different source redshifts are exactly one on top of the other, 
suggesting that the third \textit{standardized} moment is only due to geometry -- underdensities occupy more volume than overdensities -- and does not depend on the lensing efficiency. Indeed, from Eqs.~\eqref{varkappa}-\eqref{skewkappa}, considering the contribution from the redshift bin $i$, we have that:
\begin{align}
    \tilde \mu_3^i =& \frac{\sum_u \kappa_{1iu}^{3} \Delta N_{iu}}{\big (\sum_u \kappa_{1iu}^{2} \Delta N_{iu} \big)^{3/2}}
    =\frac{\sum_u \Sigma_{iu}^{3} \Delta N_{iu}}{\big (\sum_u \Sigma_{iu}^{2} \Delta N_{iu} \big)^{3/2}} \;,
\end{align}
where, after using used Eq.~\eqref{kappa1}, the lensing efficiency of Eq.~\eqref{opw} cancels out.

\subsection{Moments of the magnitude PDF}\label{momentsPDF}

We now focus on analyzing the second, third, and fourth central moments of the shift in magnitude $\Delta m$, which directly relates to magnification $\mu$ through
\begin{equation}
    \Delta m \equiv - 2.5 \log_{10} \mu.
\end{equation}
Using magnitudes is advantageous because the magnitude PDF exhibits more stable central moments than the $\mu$--PDF, which has a divergent variance in cosmology. Nevertheless, truncating the magnitude PDF at a certain negative $\Delta m$ value is prudent, as highly magnified objects could, in principle, be individually detected and excluded from the statistics either \textit{a posteriori} or during target selection \textit{a priori}.\footnote{These methods accurately recover the mean of the non-truncated PDFs, essential for unbiased cosmological analyses. Since the mean is a non-central moment, we will not discuss the effects of truncation here.}
Furthermore, as previously demonstrated, the baryonic contribution becomes dominant for $\mu \gtrsim 2.5$, which is outside the focus of this study.
In other words, we use $\Delta m_{\rm cut}$ as a proxy for a series of systematic and sample-size effects on the observable PDF.

\begin{figure}
    \centering
    \includegraphics[width=.92\linewidth]{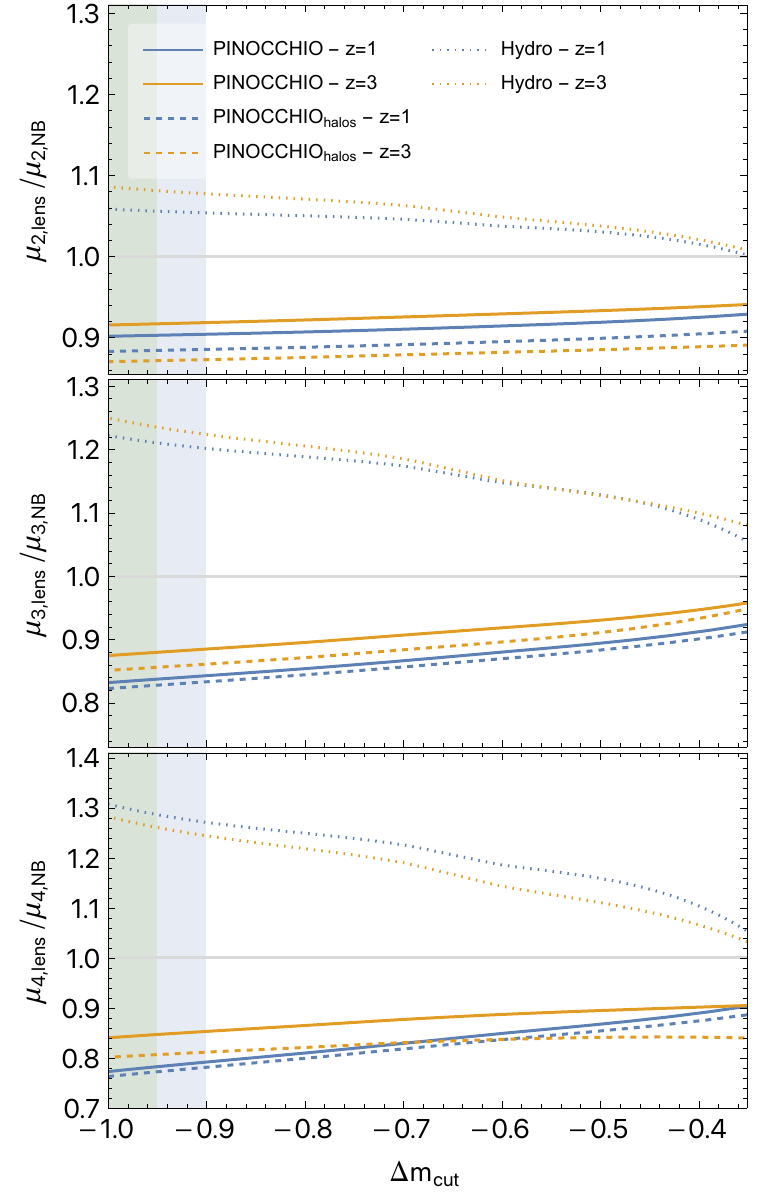}
    \caption{Central moments' ratio of the magnitude PDFs of the various methods with respect to the $N$-body simulation, for different values of $\Delta m_{\rm cut}$ and for $z = 1$ and $z= 3$. The blue (green) shaded regions indicate the limit where the baryonic effects are dominant, above 50\%, for $z=1$ ($z=3$).
    }
    \label{fig:ratio_momenta}
    \medskip
\end{figure}

Results for redshifts 1 and 3, computed in the source plane with a fixed angular resolution of 1.76 arcsec, are depicted in Fig.~\ref{fig:ratio_momenta}. These PDFs are assessed up to a cutoff magnitude, $\Delta m_{\rm cut}$, beyond which the PDF is considered zero. The second central moment shows the strongest agreement, indicating \pinocchio's capability to accurately reproduce the body rather than the tail of the PDF distribution -- a consistency noted in earlier sections. Particularly, at $z=3$, \pinocchio\ and \pinocchio$_{\rm{halos}}$ agree more closely with $N$-body results compared to $z=1$, except for $\mu_{2,\rm{lens}}$ in \pinocchio$_{\rm{halos}}$. We interpret the latter disagreement as due to the higher relative importance of non-halo particles at higher redshifts, at which halos give a less complete description of the matter field (see next section). This indicates that non-halo particles significantly impact the central moments, suggesting that the discrepancies with the $N$-body results are not simply due to the halo profiles and positioning.

\begin{figure*}
    \centering
    \subfigure{\includegraphics[width=.82\linewidth]{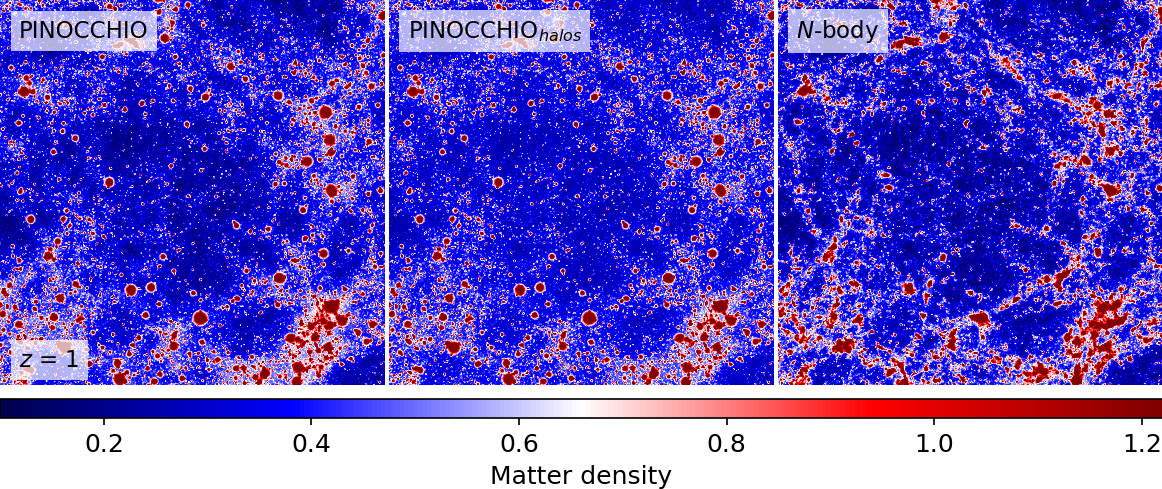}}\\
    \subfigure{\includegraphics[width=.82\linewidth]{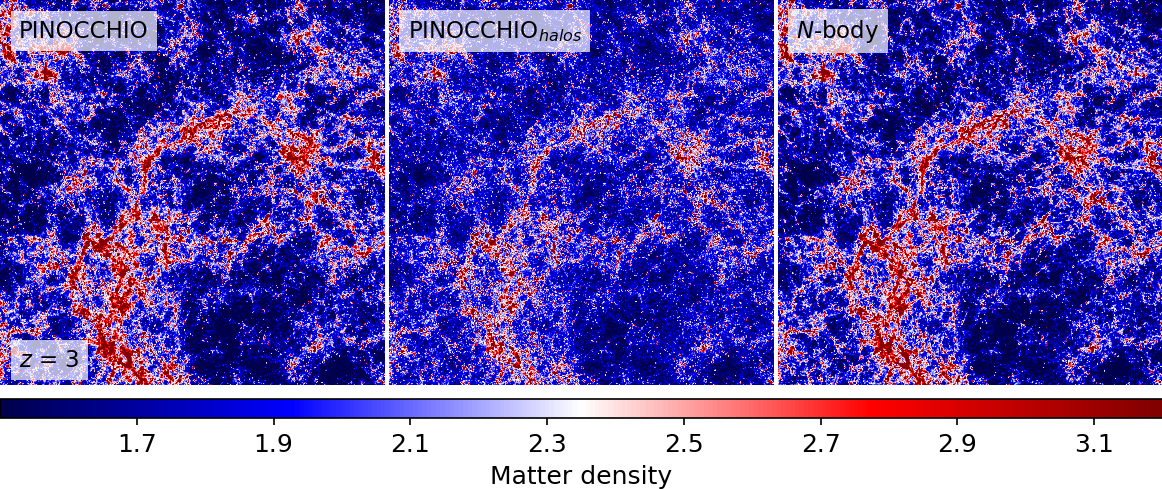}} \\
    \caption{Projected density maps for the angular resolution of 1.76 arcsec. The projected slice has a thickness of 150 Mpc/$h$. Neglecting non-collapsed matter in \pinocchio$_{\rm halos}$ leads to larger differences at higher redshifts. The differences between \pinocchio\ and $N$-body are instead similar for both redshifts. Compare with Table~\ref{tab:L1norm}.}
    \label{fig:densitymaps}
    \medskip
\end{figure*}

Fig.~\ref{fig:ratio_momenta} is also useful to quantify the impact of baryons in a way that is complementary to the ${\cal{L}}_1$ norm of Table~\ref{tab:L1norm}.
We see that for $\Delta m_{\rm cut}\simeq -0.4$, equivalent to $\mu \approx 1.5$, baryonic effects are negligible for the second moment and below 10\% for the third and fourth moments. Furthermore, we can see that the 50\% thresholds that we introduced in Table~\ref{tab:DM-ranges} correspond to 7\%, 22\% and 26\% error for the second, third and fourth moments, respectively.

Overall, \pinocchio\ well models statistical quantities within limits where baryonic effects are negligible. This precision makes it useful for constructing simple models of central moments as functions of redshift and cosmological parameters. These models are crucial for the MeMo method \citep{Quartin2014}, which utilizes the weak lensing effect on the intrinsic distribution of magnitudes to constrain key cosmological parameters, such as the amplitude of the matter power spectrum and the matter density parameter \citep{Quartin2014, Castro:2014oja,  Castro:2016jmw}.

\subsection{Maps and power spectrum}
\label{sec:maps}

Finally, to gain a deeper understanding of our findings, we will analyze the maps of the lensing quantities that underpin the PDFs discussed in previous sections, along with the convergence power spectra generated by the \pinocchio\ code and the $N$-body simulations across various redshifts. 

In Fig.~\ref{fig:densitymaps}, we display the projected density maps of matter distributed over the lens plane at redshifts $z=1$ and $z=3$, reconstructed using the $N$-body simulations and \pinocchio\ in its two configurations. These maps highlight the distribution of halos and filaments, as well as the presence of galaxy clusters at the intersections of filaments.
We observe a better agreement between \pinocchio\ and $N$-body at $z=3$, with \pinocchio$_{\rm halos}$ showing weaker structures. This is consistent with our previous observation that non-halo particles become more significant at higher redshifts, where the simulated halos provide a less comprehensive depiction of the matter field.
Indeed, the absence of halos smaller than those defined by the mass resolution of the simulation leads to a relatively greater importance of diffuse matter between halos, responsible for power on large-scale modes crucial for replicating the linear distribution of structures.
Conversely, while \pinocchio\ and \pinocchio$_{\rm halos}$ agree well at $z=1$, their correspondence with the $N$-body simulation worsens, as halo profiles and positions do not accurately replicate the small-scale patterns observed in the $N$-body simulations. These results also highlight the limitations of the LPT approximation up to the 3rd order, employed in the \pinocchio\ code, in capturing the distribution of matter on smaller scales. Notably, the absence of more massive structures, which are evident in the $N$-body simulation maps, confirms this limitation, aligning with findings from early works on the \pinocchio\ code \citep{Monaco:2016pys, Munari2017b}.

\begin{figure*}
    \centering
    \subfigure{\includegraphics[width=.93\linewidth]{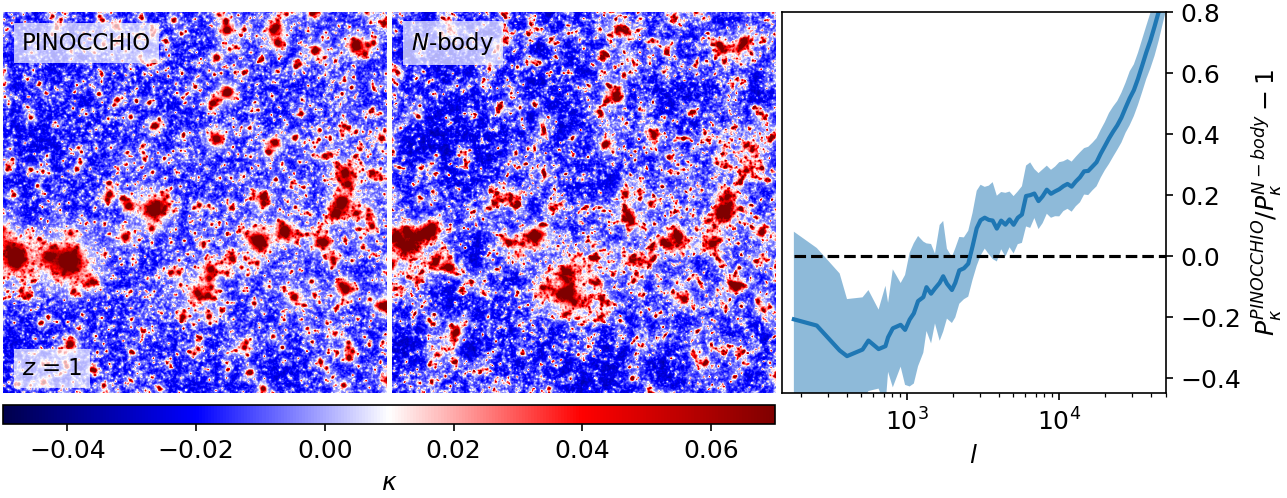}}\\
    \subfigure{\includegraphics[width=.93\linewidth]{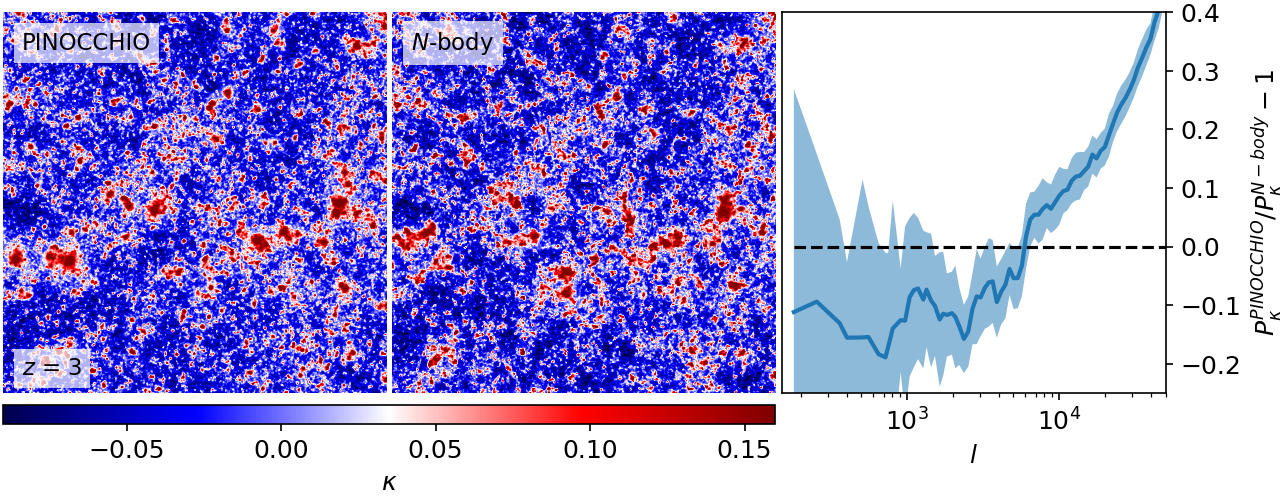}} \\
    \caption{Comparison of convergence maps and angular power spectra for the angular resolution of 1.76 arcsec. The displayed ratio represents the average across 20 PLCs, with error bands reflecting the propagated uncertainties of this average.}
    \label{fig:kappamaps}
    \medskip
\end{figure*}

\begin{figure*}
    \centering
    \subfigure{\includegraphics[width=.82\linewidth]{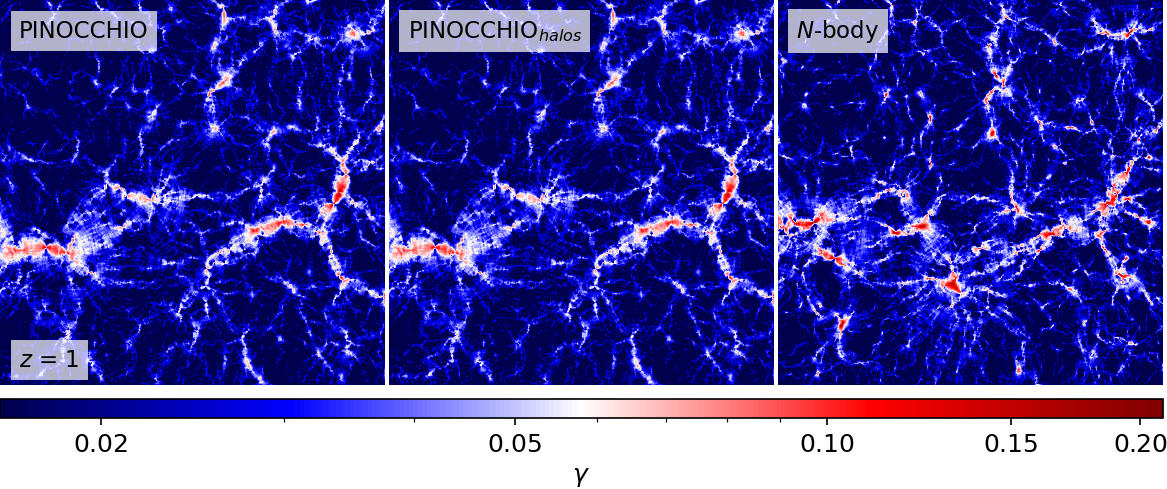}}\\
    \subfigure{\includegraphics[width=.82\linewidth]{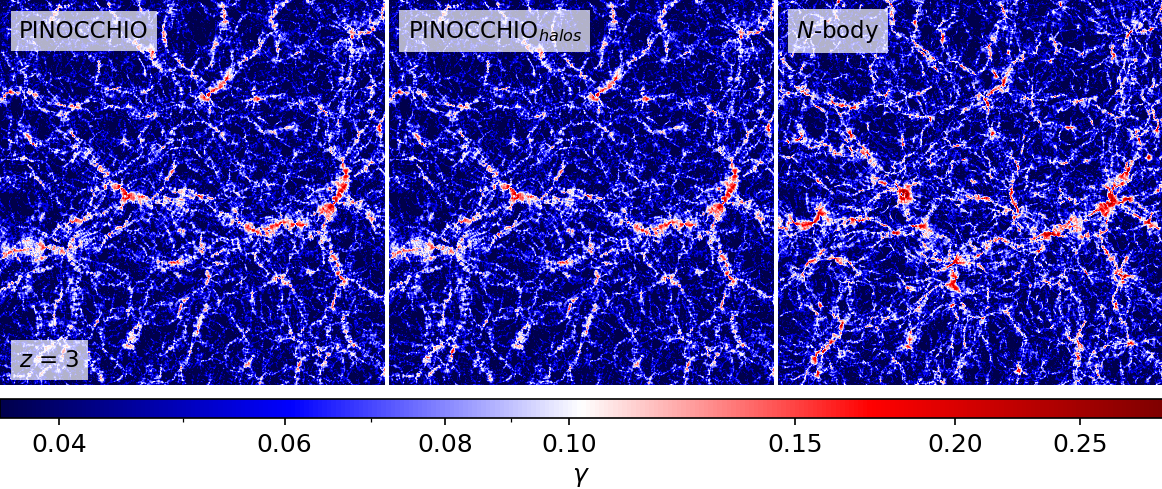}} \\
    \caption{Shear maps for the angular resolution of 1.76 arcsec. 
    The differences between \pinocchio\ and $N$-body are visible and similar for both redshifts. \pinocchio$_{\rm halos}$ produces very similar structure to \pinocchio\, and the differences are mostly on the one-point distribution and thus harder to spot.
    Compare with Table~\ref{tab:L1norm}.
    }
    \label{fig:gammamaps}
    \bigskip
\end{figure*}

The convergence maps and power spectra for the $N$-body and \pinocchio\ simulations are displayed in Fig.~\ref{fig:kappamaps}. The \pinocchio\ maps begin to diverge from those of the $N$-body simulations at small scales, yet show good agreement on larger scales. This observation is corroborated both through visual inspection and by analyzing the angular convergence power spectrum, which shows a 1--2$\sigma$ consistency for $l \lesssim 10^4$, 
consistent with the observations from Fig.~\ref{fig:densitymaps}. At small scales, the contribution of the halo profile becomes dominant, where the lack of substructures in the profile adopted here implies the disagreement observed in this limit.\footnote{See \citet{Giocoli:2017nnq} for the effects of substructures in the power spectrum and PDFs.}
Additionally, Fig.~\ref{fig:gammamaps} presents a comparison of the shear maps. There is broad agreement between the maps for both $z=1$ and $z=3$. Distinctions between \pinocchio, \pinocchio$_{\rm halos}$, and the $N$-body simulations become noticeable only at high-shear levels. Here, the filaments represented in \pinocchio\ appear more pronounced than those in the $N$-body simulation, highlighting the limits of agreement. This finding is consistent with the analysis of the PDFs detailed in Sections~\ref{sec:pdfsec} and \ref{app:stronglens}, which highlighted disagreement for $\gamma \gtrsim 0.1$. In particular, we can infer that the correct modeling of filaments is important for type II and III images.

\bigskip

\section{Conclusions}
\label{sec:concl}

In this paper, we investigated the effectiveness of the approximate methods \pinocchio\ and \texttt{turboGL} in accurately reproducing lensing probability density functions. Our simulations assume a universe composed solely of dark matter, excluding baryonic effects. We studied the impact of gravitational lensing in both weak and strong regimes, differentiating the contributions from the three types of images. We thoroughly compared the results of these approximate methods with those obtained from $N$-body simulations. Key findings include:

\begin{enumerate}

\item Approximate methods are effective primarily within the weak and medium-lensing region, which mainly arises from mildly nonlinear matter structures. Accurately modeling the tails of the PDFs necessitates $N$-body simulations, as demonstrated by our analysis of Types I, II, and III  lensing PDFs.

\item For supernovae and standard sirens, these methods prove useful for statistical analyses, given that strong lensing events are either rare or can be excluded. Within this regime, baryonic effects are expected to have a minor impact.

\item These approximate methods struggle to capture the small-scale nonlinear matter field accurately. As the perturbation theory fails to accurately predict small-scale halo clustering and intertwining matter distribution, enhancing the modeling of halo internal structures could improve results. Recent work by~\citet{Berner:2021hmp} has initiated efforts to reconstruct halo substructure from merger histories, which could be integrated into these methods to add more small-scale power to the simulations.
We also observed that non-halo particles minimally impact lensing for $z<3$.

\item Our evaluation of the second-to-fourth moments of the lensing PDFs indicated that \pinocchio\ agrees with $N$-body results to within 10\% in the regime originally proposed by the MeMo method by~\citealt{Quartin2014}, that is, $\Delta m_{\rm cut} \gtrsim -0.4$, suiting it to be used for supernovae lensing. However, it underestimates the second-to-fourth moments by roughly 10, 15, and 20\% in a regime where the baryonic effect is yet not dominant, namely $\Delta m_{\rm cut} \gtrsim -1$.

\item The stochastic simulations of \texttt{turboGL} are suitable for rapid likelihood exploration due to their speed and flexibility in handling low-statistics PDFs ($\sim10^5$ rays). In contrast, ray tracing is more effective at exploring high-$\mu$ tails as the number of rays increases with the square of the field of view, while stochastic methods scale linearly with sample size.

\item The analyses of the moments in Figs.~\ref{fig:diss-M} and~\ref{fig:diss-z} and the maps (Figs.~\ref{fig:densitymaps}--\ref{fig:gammamaps}) underscore the need for precise modeling of mass distribution and non-linear clustering of halos with $M > 10^{12} M_\odot$. The challenges are twofold: ($i$) while there is a moderate consensus on the impact of baryons at the scale of galaxy clusters, their effects in individual galaxies and galaxy groups remain far less understood and more uncertain~\citep[see, e.g.,][]{Euclid:2023jih}; ($ii$) lighter objects in filaments require both accurate modeling of their individual profiles and a detailed understanding of their interconnected mass distributions.

\end{enumerate}

These insights are expected to guide future studies on gravitational lensing of point sources, a topic of growing relevance in light of upcoming supernova and gravitational wave datasets.

\section*{Acknowledgements}

{\small
It is a pleasure to thank Ben Metcalf for the support with \glamer\ software. TC is supported by the Agenzia Spaziale Italiana (ASI) under - Euclid-FASE D  Attivita' scientifica per la missione - Accordo attuativo ASI-INAF n. 2018-23-HH.0, by the PRIN 2022 PNRR project "Space-based cosmology with Euclid: the role of High-Performance Computing" (code no. P202259YAF), by the Italian Research Center on High-Performance Computing Big Data and Quantum Computing (ICSC), project funded by European Union - NextGenerationEU - and National Recovery and Resilience Plan (NRRP) - Mission 4 Component 2, within the activities of Spoke 3, Astrophysics and Cosmos Observations, by the INFN INDARK PD51 grant, and by the FARE MIUR grant `ClustersXEuclid' R165SBKTMA.
VM is supported by the Brazilian research agencies FAPES and Conselho Nacional de Desenvolvimento Científico e Tecnológico (CNPq).
MQ is supported by the research agencies Fundação Carlos Chagas Filho de Amparo à Pesquisa do Estado do Rio de Janeiro (FAPERJ) project E-26/201.237/2022 and CNPq.
We  acknowledge the use of the Santos Dumont supercomputer of the National Laboratory of Scientific Computing (LNCC, Brazil) through the projects \texttt{lenssims} and \texttt{lensingemu}.
}

\section*{Data and Software Availability}

Codes will be shared on reasonable request to the corresponding author. Data products and post-processing routines are available at \citet{tiago_castro_2024_11093020}.\footnote{\href{https://github.com/TiagoBsCastro/LensingMethodsDeconstruction}{github.com/TiagoBsCastro/LensingMethodsDeconstruction}}



\bibliography{references}


\appendix

\begin{figure*}
    \includegraphics[width=.96\linewidth,trim=0 38 0 0 ,clip] {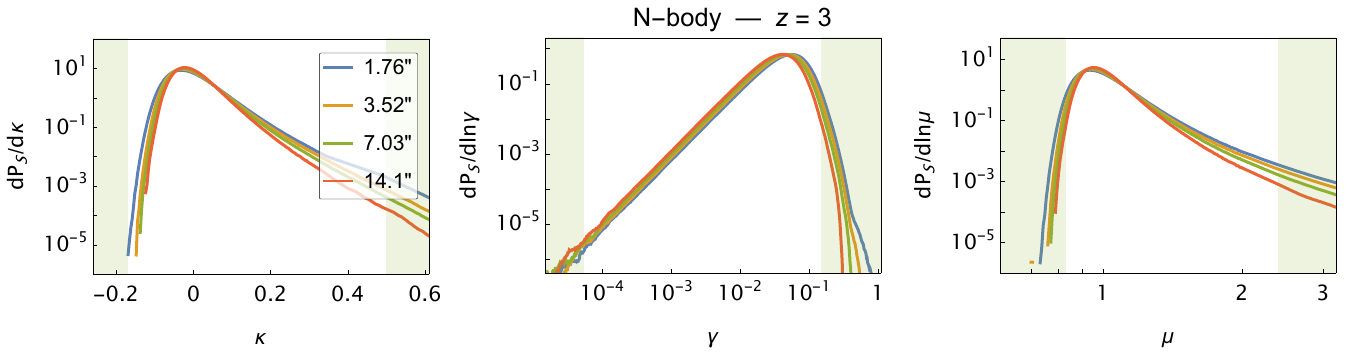}\\
    \includegraphics[width=.96\linewidth,trim=0 0 0 0 ,clip] {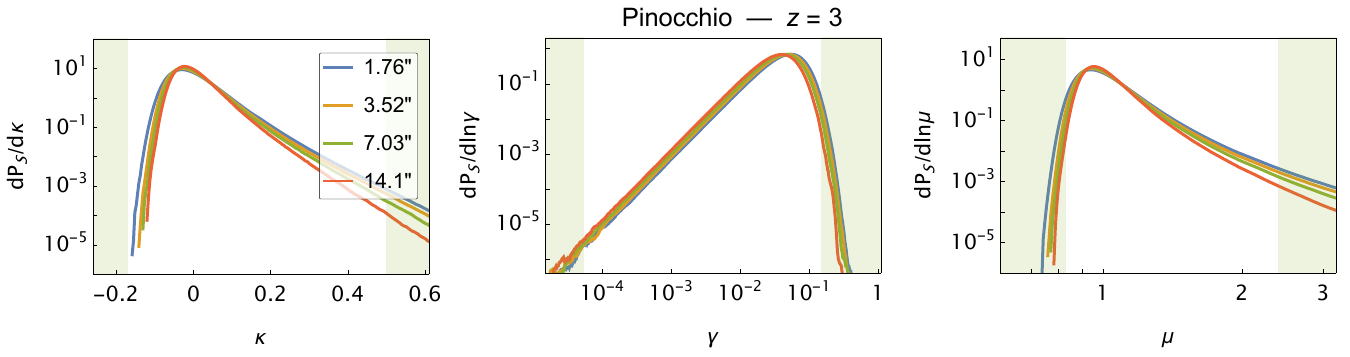}
    \caption{Similar to Figure~\ref{fig:pdf-comparison-zs}, but for different angular resolutions at $z=3$. We use the 1.76" resolution as reference for the shaded regions. \label{fig:pdf-comparison-arcsec}}
\end{figure*}

\section{PDFs at different angular resolutions}
\label{app:resolutions}

\vspace{-.1cm}

Figure~\ref{fig:pdf-comparison-arcsec} illustrates how changes in angular resolution affect the lensing PDFs from \pinocchio\ and $N$-body simulations. Increasing the angular resolution angle alters the peak and variance of the lensing PDFs in the following ways:
    \begin{enumerate}[label=\roman*)]
    \item the peak in the convergence PDF shifts to lower $|\kappa|$ values and the variance decreases;
    \item the peak in the shear PDF shifts to lower $\gamma$ values without a noticeable change in its variance;
    \item the magnification exhibits a combined effect of i) and ii), where the PDF peak shifts towards $\mu=1$ and the variance decreases.
\end{enumerate}
This behavior is anticipated as using a larger resolution angle effectively smooths the matter field, thus reducing the overall lensing effect.
Notably, the characteristic elbow at $\gamma \gtrsim 0.1$ vanishes with a larger resolution angle. As discussed in Section~\ref{app:stronglens}, this feature is linked to type II and III images, which are suppressed by the effective smoothing of the matter field.

\label{lastpage}

\end{document}